\documentclass[MSNbibl,nameyear,dvips]{arxstspdf}
\usepackage{flushend}
\usepackage{stfloats}
\usepackage{graphicx}
\volume{28}
\issue{2}
\pubyear{2013}
\firstpage{223}
\lastpage{237}
\doi{10.1214/12-STS408} 

\makeatletter
\newcommand{\rrvert}{\vert}
\newcommand{\llvert}{\vert}

\makeatother

\begin{document}
\begin{frontmatter}

\title{Mean--Variance and Expected Utility:
The~Borch Paradox}
\runtitle{Mean--Variance and Expected Utility}

\begin{aug}
\author[a]{\fnms{David} \snm{Johnstone}\corref{}\ead[label=e1]{d.johnstone@econ.usyd.edu.au}}
\and
\author[b]{\fnms{Dennis} \snm{Lindley}}
\runauthor{D. Johnstone and D. Lindley}

\affiliation{University of Sydney}

\address[a]{David Johnstone is National Australia Bank Professor of Finance,
Sydney University Business School,
University of Sydney, NSW 2006, Australia \printead{e1}.}
\address[b]{Dennis Lindley is Professor Emeritus of Statistics, ``Woodstock,'' Quay Lane,
Minehead, Somerset, TA24 5QU, United
Kingdom.}

\end{aug}

%
\begin{abstract}
The model of rational decision-making in most of economics and
statistics is
expected utility theory (EU) axiomatised by von Neumann and Morgenstern,
Savage and others. This is less the case, however, in financial
economics and
mathematical finance, where investment decisions are commonly based on the
methods of mean--variance (MV) introduced in the 1950s by Markowitz. Under
the MV framework, each available investment opportunity (``asset'') or
portfolio is represented in just two dimensions by the ex ante mean and
standard deviation $(\mu,\sigma)$ of the financial return anticipated from
that investment. Utility adherents consider that in general MV methods are
logically incoherent. Most famously, Norwegian insurance theorist Borch
presented a proof suggesting that two-dimensional MV indifference
curves cannot represent the preferences of a rational investor (he claimed that MV
indifference curves ``do not exist''). This is known as Borch's paradox and
gave rise to an important but generally little-known philosophical
literature relating MV to EU. We examine the main early contributions to
this literature, focussing on Borch's logic and the arguments by which it
has been set aside.\vspace*{12pt}
\end{abstract}

%
\begin{keyword}
\kwd{Mean--variance}
\kwd{expected utility}
\kwd{Borch's paradox}
\kwd{probability mixture}
\kwd{portfolio theory}
\kwd{CAPM}
\end{keyword}\vspace*{12pt}

\end{frontmatter}

\begin{quotation}
There is no inevitable connection between the validity of the
expected utility maxim and the validity of portfolio analysis based on, say,
expected return and variance (Mar\-kowitz, \citeyear{Mar59}, page 209).
\end{quotation}

\section{Introduction}

This paper looks back at a little-known but highly interesting chapter in
the history of business decision-making (call it ``investment'') under
uncertainty. In a once f\^{e}ted but now rarely mentioned paper, titled
politely \textit{A Note on Uncertainty and Indifference Curves}, the
Norwegian insurance theorist and economist Karl \citet{Bor69} argued
that the
mean--variance theory of investment, invented and popularized by Markowitz
(\citeyear{Mar52}, \citeyear{Mar59}), is logically absurd. In this delightfully provocative note,
\citet{Bor69} proved, he claimed, that it is impossible to draw indifference
curves in the mean--variance $(\mu,\sigma^{2})$ or
mean--standard deviation $%
(\mu,\sigma)$ plane. The same proof appears in at least two other
works by
Borch (\citeyear{Bor73}, \citeyear{Bor74}), who concluded that mean--variance is an interesting but
not serious alternative to expected utility:

\begin{quotation}
$\ldots$ I shall continue to use mean--variance analysis in teaching, but I shall
warn students that such analysis must not be taken seriously and
applied in
practice (\cite{Bor74}, page 430).
\end{quotation}

The proof presented by Borch (pronounced\break ``Bork'') became known to theorists
as ``Borch's paradox''. While of much interest theoretically, the academic
discussion that stemmed from Borch's work had virtually no impact on the
practice of finance. To the contrary, mean--variance (MV) analysis, for which
Markowitz later won a Nobel Prize in Economic Sciences, has become by far
the most recognized decision framework in the practice of business
decision-making, including especially capital budgeting (e.g., whether
to build a new
factory), investment management (e.g., whether to increase the weight
of oil
stocks in a pension fund) and corporate financial valuation (e.g.,
whether a
firm is worth its current value on the stock market). Each of these common
applications is built implicitly on MV, and explicitly on the so-called
``capital asset pricing model'' (CAPM) that arose as a corollary from the MV
foundations set out by Markowitz.

Although business applications of MV portfolio theory and the CAPM are
commonplace, and effectively the industry standard (witness any modern
textbook in financial economics), proponents of this decision framework
remain conscious that the proven philosophical foundations of decision
analysis under uncertainty remain the axioms and theorems of expected
utility theory (EU), formalized by \citet{vonMor53} and
\citet{Sav54}. Utility theory, or more specifically the
maximization of subjective expected utility satisfying the von
Neumann--Morgenstern or similar axioms, remains the hallmark of
rationality in economics and statistical decision analysis (e.g.,
\cite{DeG70}; \cite{BerSmi94}; \cite{PraRaiSch95};
\cite{Len04}; \cite{EecGolSch05}). As a mark of respect for this
intellectual legacy, \citet{Mar91} devoted his 1990 Nobel Lecture
to an empirical comparison of his MV methods of portfolio \mbox{selection}
with a model based on EU theory. Authoritative recognition also goes
the other way. In its second edition, one of the standard references on
neoclassical decision theory, \citet{PraRaiSch95} contains an elegant
exposition of the MV investment framework, albeit without
reconciliation with other explicitly EU parts of the book. Completing
the circle, \citet{Mar} and \citet{Rub06N1} have lately
pointed to an early paper in Italian, authored by \citet{deF40},
the most revered of all subjective probability theorists, as having
been first to express a formal model of decision-making within a MV
framework. Two further expositions concentrated on de Finetti's
previously little-known anticipation of Markowitz are
\citet{Bar08} and \citet{PreSer07}.

Our primary purpose is to examine the historical literature surrounding
Borch's paradox. To assist readers who are not familiar with this
branch of
applied statistical literature, we first recount the basic elements of
investment decision-making under the two competing conceptual frameworks,
expected utility and mean--variance. We then consider how MV can be justified
on axiomatic foundations, in the face of critics such as Borch, and by
comparison with EU theory generally. Finally, to better understand the
practical appeal of MV methods, and why finance theory so readily adopted
the language of MV over EU, we introduce the capital asset pricing model
(CAPM) and observe how such a theoretically insightful model arose almost
automatically once decisions were depicted in terms of MV rather than EU.

\section{Expected Utility Theory}

A decision is a choice between some (usually strict) subset of all of
the available ``lotteries,'' ``assets,'' ``investments'' (these terms
are synonyms)---and all feasible weighted portfolios thereof. Each such
uncertain prospect reduces to a probability distribution over a domain
of possible payoffs. Decision-making is therefore boiled down to a
choice between different possible probability distributions of returns.

\citet{vonMor53} proposed an axiomatic theory of how to
decide between known probability distributions (of payoffs). In brief, they
proved deductively that if decision-making is logical in the sense that it
obeys certain specified basic axioms of coherence or rationality, then
implicitly the decision-maker must act \textit{as if} her objective is to
maximize expected utility $E[u(x)]=\int_{x}f(x)u(x)$, where $u(x)$
is a real-valued function representing the utility obtained from certain
wealth or payoff $x$, and $f(x)$ is the probability density function of $x$.
The decision rule of maximizing $E[u(x)]$, taken in conjunction with some
plausible looking utility function such as Bernoulli's $u(x)=\log(x)$, is
often treated as itself axiomatic. More correctly, the extra-intuitive
appeal of the EU decision rule is that rather than being just another
plausible looking but arbitrary objective function, it is a theorem deduced
from a small number of far more elementary assumptions concerning what
constitutes rational human preferences.

There are five essential axioms of expected utility: (i)
\textit{Completeness}. All lotteries $A$ and $B$ can be ranked relative
to one another, $A\succ B $, $A\prec B$ or $A\sim B$, where $\sim$
indicates indifference. (ii) \textit{Transitivity}. If $A\succeq B$ and
$B\succeq C$, then $A\succeq C$. (iii) \textit{Continuity}. If
$A\succeq B\succeq C$, there exists some probability
$\alpha\in\lbrack0,1]$ such that $B\sim\alpha
A+(1-\alpha)C$%
, meaning that $B$ is indifferent to a compound lottery that returns lottery
$A$ with probability $\alpha$ and lottery $C$ with probability
$(1-\alpha)$. (iv) \textit{Independence}. Indifference $A\sim B$ between
lotteries $A$
and $B$ implies indifference between compound lottery $\alpha
A+(1-\alpha
)C $ and compound lottery $\alpha B+(1-\alpha)C$. Similarly, $A\succ B$
implies $\alpha A+(1-\alpha)C\succ\alpha B+(1-\alpha)C$, for $\alpha>0$.
(v) \textit{Dominance}. Let $C_{1}$ be the compound lottery $\alpha
_{1}A+(1-\alpha_{1})B$ and let $C_{2}$ be the compound lottery $\alpha
_{2}A+(1-\alpha_{2})B$. If $A\succ B$, then $C_{1}\succ C_{2}$
if and
only if $\alpha_{1}>\alpha_{2}$.

For further interpretation of these axioms and\break proof of how they lead
to the
von Neumann and Morgenstern EU decision rule, see Pennacchi (\citeyear{Pen08}),
pages 4--11.
Similar expositions are found in many textbooks both in economics and
statistics. See, for example, Ingersoll (\citeyear{Ing}), pages 30--44, Huang and
Litzenberger (\citeyear{HuaLit88}), pages 1--11, and
Levy (\citeyear{Lev12}), pages 25--30. In brief, by
assuming the primitive preference relationships \mbox{(i)--(v)}, it is shown that
lottery $A$ is preferred to lottery $B$ if and only if the expected utility
of lottery $A$ exceeds the expected utility of lottery $B$. Expected utility
$E[u(x)]$ is therefore the proven measure by which to rank uncertain
investments.

The usual assumption in economics is that decision-makers are risk averse.
This means that they have positive but diminishing marginal utility for
money, and hence $u(x)$ is increasing and concave. A risk averse
decision-maker will not accept any lottery with an expected money value
of zero (or
less). Actuarially fair bets are thus unacceptable. Before accepting a bet
to win or lose some fixed sum $c$, a risk averse agent requires that the
probability of winning exceeds 0.5 by some premium. The amount of this
premium depends on the local concavity of $u(x)$ or on how fast the marginal
utility of money is diminishing in the region of wealth $x_{0}\pm c$,
where $%
x_{0}$ is her starting wealth. Technically, the Pratt--Arrow measure of local
absolute risk aversion $-u^{\prime\prime}(x)/u^{\prime}(x)$ captures the
degree of concavity of $u$ or the rate at which marginal utility is
decreasing at wealth $x$.

\section{Mean--Variance Theory}

The following quick summary of MV owes much to \citet{Liu04}. The one-period
return on an investment over period $t$ is defined as $(p_{t}+d)/p_{t-1}$,
where $p_{t}$ is the time $t$ asset price and $d$ is the income (dividend)
drawn from the asset in period $t$. This definition has the advantage that
the returns measure is always positive.

\begin{figure}[b]

\includegraphics{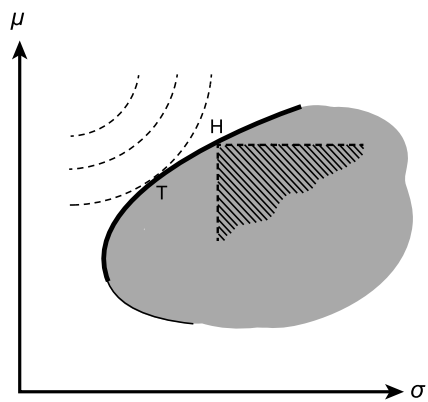}

\caption{Graphical presentation of $\operatorname{MV} (\sigma,\mu)$ analysis.}\label{fig1}
\end{figure}

Imagine a set of available investments or ``assets''. These can be combined
into arbitrarily weighted portfolios (e.g., the investor might form a 2:1
weight\-ed portfolio of assets A and B, where two-thirds of her money is
invested in A and one-third in B). The available assets and their linearly
weighted portfolios form an opportunity set of investments. Each possible
asset or portfolio presents a compromise between mean return $\mu$ and
variance $\sigma^{2}$. Each such MV pair is reduced following
Markowitz and
finance convention to its parameters $(\sigma,\mu)$. The opportunity set
is then a region of feasible $(\sigma,\mu)$ pairs. This region is depicted
in its characteristic shape by the bullet-like shaded area in Figure~\ref{fig1}. The
investor is generally risk averse and thus prefers portfolios with higher
mean return $\mu$ and lower ``risk'' (standard deviation) $\sigma$. The
opportunity set is reduced therefore to just those portfolios on the thick
black arc called the ``efficient frontier''. Each asset portfolio on the
efficient frontier dominates all assets and portfolios to its southeast,
because these have both lower $\mu$ and higher $\sigma$. For example,
asset $(\sigma,\mu)$-pair $H$ dominates all assets and portfolios in the
hatched region.\looseness=1\vadjust{\goodbreak}

To choose between all efficient portfolios, the investor forms a family of
MV indifference curves. These are understood as equivalue curves
defined by
some indifference function $V(\sigma,\mu)$. Each such curve shows the
locus of $(\sigma,\mu)$ points for which $V(\sigma,\mu)$ is held
constant. A typical looking family of indifference curves is shown by the
dotted lines in Figure~\ref{fig1}. These curves are drawn convex downwards on the
basis that, for assets known only by their MV parameters $(\sigma,\mu
)$, the risk averse investor requires marginally greater compensation in
$\mu$ for each further increment in risk $\sigma$ (see \cite{Mey87}, for related
proofs). Since risk averse investors prefer lower $\sigma$ for fixed
$\mu$%
, higher (more northwesterly) indifference curves represent greater
expected utility to the investor. Having established both the efficient
frontier and an indifference function $V(\sigma,\mu)$, the unique
MV-optimal investment in risky assets is located at the point of tangency $T$ between the
decision-maker's own $V(\sigma,\mu)$ and the exogenous efficient frontier.

\section{Borch's Paradox}

Borch (\citeyear{Bor69}), pages 2 and 3, presented a proof based on assets with two-point
distributions that revealed (he claimed) that it is impossible to draw
indifference curves in the mean--variance $(\mu,\sigma^{2})$ or
mean--standard deviation $(\mu,\sigma)$ plane. Borch repeat\-ed this same
proof in 1973 and 1974, and wrote openly of his frustration that it was
not more widely acknowledged by theorists developing portfolio optimization
methods:

\begin{quotation}
I have on several occasions (1969) and (1974) tried to warn against the
uncritical use of mean--variance analysis. It was probably too much to expect
that these warnings should have much effect (\cite{Bor78}, page~181).
\end{quotation}

The Borch paradox goes as follows. First, assume that two assets (payoff
distributions) with parameters $(\mu_{1},\sigma_{1}^{2})$ and $(\mu
_{2},\sigma_{2}^{2})$ are regarded, on the basis of those parameters alone,
as indifferent (i.e., of equal subjective merit). Now imagine two
hypothetical assets constructed simply as two-point distributions.
\mbox{Asset~1}
produces payoff $y_{1}$ with probability $p$ and payoff $x$ with probability
$(1-p)$. Asset 2 produces payoff $y_{2}$ with probability $p$ and
payoff $x$
with probability $(1-p)$. By common sense or some very basic axiom like the
``sure thing'' principle raised by Savage (Borch cites Allais' concept of
``preference absolue''), these two assets are indifferent if and only if $
y_{1}=y_{2}$ $(p>0)$.\vadjust{\goodbreak}

Now, suppose that the constants $x$, $p$, $y_{1}$ and $y_{2}$ take values
%
\begin{eqnarray}
\label{equ1}
x &=&\frac{\sigma_{1}\mu_{2}-\sigma_{2}\mu_{1}}{\sigma_{1}-\sigma_{2}},
\\
\label{equ2}
p &=&\frac{(\mu_{1}-\mu_{2})^{2}}{(\mu_{1}-\mu_{2})^{2}+(\sigma
_{1}-\sigma_{2})^{2}},
\\
\label{equ3}
y_{1} &=&\mu_{1}+\sigma_{1}\frac{(\sigma_{1}-\sigma_{2})}{(\mu
_{1}-\mu
_{2})},
\\
\label{equ4}
y_{2} &=&\mu_{2}+\sigma_{2}\frac{(\sigma_{1}-\sigma_{2})}{(\mu
_{1}-\mu
_{2})}.
\end{eqnarray}
Borch did not make any mention of where these equations come from or what
they assume, except to say that it is easy to verify that the implied values
of the mean and variance parameters of the two assets are,
respectively,
$(\mu
_{1},\sigma_{1}^{2})$ and $(\mu_{2},\sigma_{2}^{2})$, thus matching
assets~1 and 2 (this is indeed easily verified).

The final step in Borch's proof holds that because the two assets can
be of equal merit only if $y_{1}=y_{2}$, the indifference condition
$\mbox{(\ref{equ3})}=\mbox{(\ref{equ4})}$ is
\[
\mu_{1}+\sigma_{1}\frac{(\sigma_{1}-\sigma_{2})}{(\mu_{1}-\mu
_{2})}%
=
\mu_{2}+\sigma_{2}\frac{(\sigma_{1}-\sigma_{2})}{(\mu_{1}-\mu_{2})},
\]
implying that $(\mu_{1}-\mu_{2})^{2}+(\sigma_{1}-\sigma_{2})^{2}=0$ and,
hence, indifference requires that $\mu_{1}=\mu_{2}$ and $\sigma
_{1}=\sigma_{2}.$ According to Borch's interpretation of this result, any
supposed indifference between two arbitrary mean--variance pairs is
impossible, unless of course they are the same. Mean--variance indifference
curves are thus merely points rather than curves, or, in Borch's
[(\citeyear{Bor69}), page 3]
own words, ``it is impossible to draw indifference curves in the
E--S-plane''
(E and S denote the mean and standard deviation).

In answer to any suspicion raised by their nonexplanation, there is nothing
contrived about the four equations used by Borch to define the
constants $x$%
, $p$, $y_{1}$ and $y_{2}$ in his proof. Rather, these can be derived
by writing the standard equations for the means and standard
deviations, $\mu _{1}$, $\mu_{2}$, $\sigma_{1}$ and $\sigma_{2}$, of
the two Borch two-point assets in terms of $x$, $p$, $y_{1}$ and
$y_{2}$, and then solving these four equations simultaneously to get
general expressions for all four constants in terms of the specified
means and standard deviations. It follows, therefore, that the four
Borch equations, numbered (\ref{equ1})--(\ref{equ4}) above, are not
merely sufficient conditions to produce the specified mean--variance
parameters $(\mu_{1},\sigma_{1}^{2})$ and $(\mu_{2},\sigma_{2}^{2})$.
Rather, they follow necessarily from those specified parameter values
as one of two possible sets of solutions. The second set of solutions,
which Borch did not raise but\vadjust{\goodbreak} could have employed to the same effect,
is as follows:
\begin{eqnarray*}
x &=&\frac{\sigma_{1}\mu_{2}+\sigma_{2}\mu_{1}}{\sigma_{1}+\sigma_{2}},
\\
p &=&\frac{(\mu_{1}-\mu_{2})^{2}}{(\mu_{1}-\mu_{2})^{2}+(\sigma
_{1}+\sigma_{2})^{2}},
\\
y_{1} &=&\mu_{1}+\sigma_{1}\frac{(\sigma_{1}+\sigma_{2})}{(\mu
_{1}-\mu
_{2})},
\\
y_{2} &=&\mu_{2}+\sigma_{2}\frac{(\sigma_{1}+\sigma_{2})}{(\mu
_{1}-\mu
_{2})}.
\end{eqnarray*}

\subsection{Our Interpretation of Borch}

Borch interprets his paradox to say that mean--variance indifference curves
cannot exist. This is too strong, as will be seen in the sections below.
More reasonable interpretations of Borch's proof are as follows.

\textit{Interpretation} 1: Suppose that a decision-maker is adamant
that he
is indifferent between any two assets with mean--variance
characteristics $%
(\mu_{1},\sigma_{1}^{2})$ and $(\mu_{2},\sigma_{2}^{2})$, where
$(\mu
_{1},\sigma_{1}^{2})\neq(\mu_{2},\sigma_{2}^{2})$. It is possible to
construct two-point assets with these very characteristics between
which no
one can reasonably be indifferent [these can be constructed by Borch's
equations (\ref{equ1})--(\ref{equ4}), or equally well with the second possible set of solutions
noted above].

This possibility does not imply that there are no possible assets with
parameters $(\mu_{1},\sigma_{1}^{2})$ and $(\mu_{2},\sigma
_{2}^{2})$
that are rationally (to someone) indifferent. Rather, it shows that the
decision-maker cannot be indifferent between all imaginable pairs of assets
with these parameters.

\textit{Interpretation} 2: If we limit consideration to only the particular
subclass of two-point assets construct\-ed by Borch, there exists no pair of
assets with $(\mu_{1},\sigma_{1}^{2})\neq(\mu_{2},\sigma_{2}^{2})$
between which anyone might reasonably be indifferent. Rather, whenever
$(\mu
_{1},\sigma_{1}^{2})\neq(\mu_{2},\sigma_{2}^{2})$, the two assets
(having $x$ and $p$ in common) necessarily differ in that $y_{1}\neq y_{2}$,
which of itself means that they cannot be indifferent.

\subsection{Numerical Illustration}

Here we exemplify \textit{Interpretation} 1 numerically. Imagine that the
subject of the experiment feels that he is indifferent between any two
assets with parameters $(\mu_{1},\sigma_{1}^{2})=(10,225)$ and $(\mu
_{2},\sigma_{2}^{2})=(20,625)$.\break Such subjective indifference would
typically require that the security with the bigger mean has the bigger
variance, but that practicality is not necessary in Borch's demonstration.
Now consider two comparable lottery tickets, ticket A and ticket B.
Ticket A
pays $25$ with probability $p=0.5$ and $-5$ with probability $(1-p)$.
Similarly, ticket B pays $45$ with probability $p=0.5$ and $-5$ with
probability $(1-p)$. The mean--variance parameters of these two
lotteries are
$(10,225)$ and $(20,625)$, respectively. Yet contrary to any thought that
two assets with these parameters are indifferent, ticket B is obviously
preferred because it has the same probability of winning as ticket~A, and
the same payoff if it loses, but pays 45 instead of 25 when it wins. Borch
saw this apparent contradiction as proof that the decision-maker cannot
logically be indifferent between two investments by reference only to their
means and variances.\looseness=1

\section{Baron's Rebuttal of Borch}

Borch's paradox is well known to those economic theorists mindful of
foundations and interested in the history of mean--variance, yet is
largely unknown elsewhere and goes unmentioned in standard finance and
financial economics texts, even in highly sophisticated works such as
\citet{Ing}, \citet{Coc01}, \citet{Bar03}, \citet{Len04}
and Pennacchi (\citeyear{Pen08}) that deal with the connections between
mean--variance models and expected utility theory. Neither is Borch
mentioned in the very thorough historical annotated bibliography of
\citet{Rub06N2}. This omission is justified perhaps by the
findings of a similarly important but now rarely mentioned paper by
\citet{Bar77}.

Baron rebuts Borch's paradox in two steps. First comes the proposition
that decision-making based on just the two parameters, mean and
variance, implies an underlying quadratic utility function. The same
argument arises in Hanoch and Levy [(\citeyear{HanLev70}), page 182] and
Sarnat [(\citeyear{Sar74}), page 687] who both note that quadratic
(second order polynomial) utility is the only form of mathematical
utility function for which expected utility reduces to a function of
just the first two moments of the payoff distribution. Specifically,
for risk-averse quadratic utility $u(x)=2ax-x^{2}$, the expected
utility is $E [
u(x)%
] =2a\mu-(\mu^{2}+\sigma^{2})$. Similarly, see the derivation by
Liu (\citeyear{Liu04}), page 233.

The presumption that MV necessarily implies qua\-dratic utility traces to
Markowitz [(\citeyear{Mar59}), page 288] and also Mossin (\citeyear{autokey44}),
pages 26 and 27.
Hanoch and Levy
[(\citeyear{HanLev70}), page 182] hold that ``rejection of quadratic utility implies the
rejection of any analysis based on the expected utility maxim'', which is
indirectly saying that the only unconditional way of hanging\vadjust{\goodbreak} onto EU while
applying MV is to adopt quadratic utility. \citet{JohLin11} have
more recently given an elementary proof revealing that, in the absence of
any further premise, MV necessitates quadratic utility.

Having concluded that Borch's proposed subjective value function can arise
from only a quadratic utility function, Baron's second step is to show that
one of the asset pairs in Borch's counterexample, $(\mu_{1},\sigma
_{1}^{2})$ and $(\mu_{2},\sigma_{2}^{2})$, must involve a potential payoff
in the domain where quadratic utility is decreasing with money. This finding
is easy to understand intuitively. The two Borch assets are identical except
that $y_{1}\neq y_{2}$ and, hence, the asset with the lower $y$ (e.g.,
asset 1
if $y_{1}<y_{2}$) can only be as good in someone's mind as the asset with
the higher $y$ if that higher $y$ is in the domain where money has negative
marginal utility, thus allowing $y_{1}$ and $y_{2}$ to have the same utility
$u(y_{i})$.

The net effect of Baron's argument can be summarized as follows:

\begin{longlist}[(iii)]
\item[(i)] Borch's paradox proves only that for any two mean--variance pairs,
$(\mu
_{1},\sigma_{1}^{2})$ and $(\mu_{2},\sigma_{2}^{2})$, a rational
decision-maker cannot be indifferent between \textit{all} pairs of
assets with the
specified parameters. Indeed, precisely as Borch revealed, there are easily
definable assets with such characteristics which are obviously not
indifferent.

\item[(ii)] If the decision-maker has quadratic utility, EU can be written as a
function of mean and variance alone and, hence, $(\sigma,\mu)$ indifference
curves do exist. It is necessary, however, to constrain the class of assets
under consideration so as to exclude any asset with one (or more than one)
potential payoff in the region where utility decreases with money. Negative
marginal utility for some $x$ is a well-known limitation of quadratic
utility, and is bound to produce irrational or incoherent decisions even
under EU if the admissible asset class is not suitably
restricted.

\item[(iii)] If assets with possible payoffs in the domain where quadratic utility
decreases with money (i.e., where the last increment of payoff brings a
reduction in utility) are excluded a priori from
consideration, as
if they cannot exist, then the class of counterexamples constructed by
Borch (and illustrated numerically above) no longer exists.
\end{longlist}

Another somewhat forgotten finding should be\break mentioned here. Taking MV
as a
representation of quadratic utility, and constraining all possible payoffs
into the domain where quadratic utility is increasing, Levy and Sarnat
[(\citeyear{LevSar72}), pages 387 and 388] proved that one MV asset pair $(\mu_{1},\sigma_{1}^{2})$
has higher utility than another $(\mu_{2},\sigma_{2}^{2})$, with $\mu
_{1}>\mu_{2}$, if and only if $(\mu_{1}-\mu_{2})^{2}-(\sigma
_{1}-\sigma
_{2})^{2}>0.$ This is a stronger condition than the usual definition of
dominance (i.e., $\mu_{1}\geq\mu_{2}$ and $\sigma_{1}<\sigma_{2}$
or $%
\mu_{1}>\mu_{2}$ and $\sigma_{1}\leq\sigma_{2}$) and is therefore more
``efficient'' in the sense that it reduces the class of possible investments
to a smaller number.

\section{Buridan's Axiom and Mean--Variance}\label{sec6}

We now summarize our own disproof of generalized mean--variance analysis,
very much in spirit with Borch, and then side with Baron by considering
possible theoretical and practical restrictions on the admissible asset
class that allow a partial reconciliation between the two ways of
decision-making.

Following a convention in finance dating to Mar\-kowitz's original exposition
of the mean--variance framework, our analysis is set out in terms of the
standard deviation $\sigma$ rather than variance $\sigma^{2}$. Adhering
to another well-entrenched custom in the finance literature, we work
with $%
\sigma$ as abscissa and $\mu$ as ordinate.

\subsection{\texorpdfstring{Decision Axioms in Terms of $(\sigma,\mu)$}{Decision Axioms in Terms of (sigma, mu)}}

Suppose there exists a value function $g(\sigma,\mu)$ that captures the
merit or goodness of a $(\sigma,\mu)$-asset such that larger $g$ implies
greater value. Indifference between two assets $(\sigma_{1},\mu_{1})$
and $%
(\sigma_{2},\mu_{2})$ means that $g(\sigma_{1},\mu_{1})=g(\sigma
_{2},\mu_{2})$. It is not necessary to be explicit about the form of
$g$.\vspace*{9pt}

\textit{Continuity-monotonicity-finiteness} (\textit{CMF}) \textit{axiom}. The merit
function $%
g(\sigma,\mu)$ is continuous, strictly increasing in $\mu$ for
every $%
\sigma$, and strictly decreasing in $\sigma$ for every $\mu$. These
properties hold throughout the $(\sigma,\mu)$ half-plane, $\sigma
\geq0$.
Continuity implies that there is no abrupt change in merit as either
$\sigma
$ or $\mu$ changes slightly. Strict monotonicity reflects the merit, either
positive or negative, of any change in $\mu$ or $\sigma$, however small.
Finiteness requires that any finite increase in $\sigma$ can be
offset by
a sufficiently large finite increase in~$\mu$. The existence of such a
merit function implies transitivity, meaning that if asset X is
preferred to
Y, and Y to Z, then X is preferred to Z (and likewise when preference is
replaced by indifference).\vspace*{9pt}

\textit{Buridan's axiom}. If a decision-maker is indifferent between two
assets $i=1$ and $i=2$, then he must also be indifferent\vadjust{\goodbreak} between either
asset and a probability-mixture asset that yields (the same payoff as) $i=1$
with probability $\alpha$ and (the same payoff as) $i=2$ with
probability $%
(1-\alpha)$, where $\alpha$ takes any value $\alpha\in\lbrack0,1]$.
Thus, according to the merit function $g(\sigma,\mu)$, indifference $%
g(\sigma_{1},\mu_{1})=g(\sigma_{2},\mu_{2})$ implies $g(\sigma
_{1},\mu
_{1})=g(\sigma_{2},\mu_{2})=g(\sigma_{\alpha},\mu_{\alpha})$,
where $%
\sigma_{\alpha}$ and $\mu_{\alpha}$ represent the mean and standard
deviation of an $\alpha$-mixture of the two indifferent ``pure'' assets (for
any probability $\alpha$). Note that Buridan's axiom is a simple corollary
of the independence axiom in EU.

\begin{figure}

\includegraphics{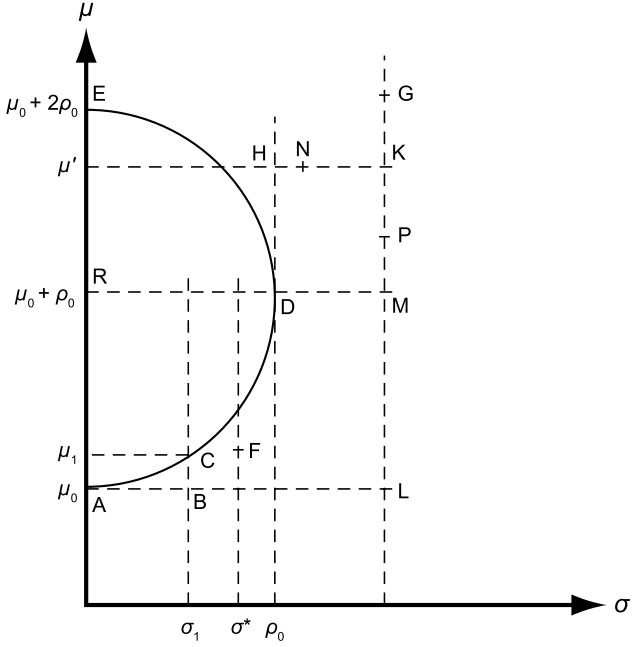}

\caption{Axiomatic decision analysis in terms of $(\sigma,\mu)$.}\label{fig2}
\end{figure}

The question now is whether a decision analysis constructed solely in terms
of $(\sigma,\mu)$ can coexist with these two axioms. Consider an asset
with $\sigma=0$, known in finance as a
``risk-free'' asset and approximated by government bonds. Let
this asset return $\mu_{0}$, as represented by the point A in Figure~\ref{fig2} with
coordinates $(0,\mu_{0})$. Now take any fixed $\sigma_{1}>0$ and
points $%
(\sigma_{1},\mu)$ for all $\mu>0$. These represent risky assets, among
which asset B with coordinates $(\sigma_{1},\mu_{0})$ is inferior to (has
lower ``merit'' than) asset A because it
has $\mu=\mu_{0}$ but greater~$\sigma$. As $\mu$ increases from
$\mu
_{0} $, the assets on $\sigma=\sigma_{1}$ increase in merit, such
that at
some point C, with coordinates $(\sigma_{1},\mu_{1})$, the higher
return $%
\mu_{1}=\mu_{0}$ is just sufficient to compensate for the associated
risk $%
\sigma_{1}$, leaving the decision-maker indifferent between C and A. The
existence of $\mu_{1}$ is guaranteed by the CMF axiom.

Assuming that the decision-maker is indifferent between asset A at
$(\sigma
_{0}=0,\mu_{0})$ and asset C at $(\sigma_{1},\break\mu_{1})$,\vadjust{\goodbreak} Buridan's axiom
dictates that this indifference extends to a randomized mixture of A
and C,
where A is selected with chance $\alpha$. The payoff $x$ from such an $
\alpha$-mixture asset of any pair $(\sigma_{0},\mu_{0})$ and
$(\sigma
_{1},\mu_{1})$ has expectations%
%
\begin{equation}
\label{equ5}
\mu=E(x)=\alpha\mu_{0}+(1-\alpha)\mu_{1}
\end{equation}
and%
\begin{eqnarray*}
E\bigl(x^{2}\bigr) &=&\alpha\bigl(\mu_{0}^{2}+
\sigma_{0}^{2}\bigr)+(1-\alpha) \bigl(\mu
_{1}^{2}+\sigma_{1}^{2}\bigr)
\\
&=&\alpha\mu_{0}^{2}+(1-\alpha) \bigl(\mu_{1}^{2}+
\sigma _{1}^{2}\bigr)\quad(\sigma_{0}=0).
\end{eqnarray*}
Since var$(x)=E(x^{2})-E(x)^{2}$, simple algebra gives%
%
\begin{eqnarray}
\label{equ6}
\sigma^{2} &=&\operatorname{var}(x)\hspace*{-10pt}\nonumber\\
&=&\alpha\sigma_{0}^{2}+(1-
\alpha )\sigma _{1}^{2}+\alpha(1-\alpha) (\mu_{1}-
\mu_{0})^{2}
\hspace*{-10pt}\\
&=&(1-\alpha)\sigma_{1}^{2}+\alpha(1-\alpha) (
\mu_{1}-\mu_{0})^{2}\quad
(\sigma_{0}=0)\hspace*{-10pt}
\nonumber
\end{eqnarray}
in the same way as found by Baron (\citeyear{Bar77}), page 1685. Note that
equations (\ref{equ5})
and (\ref{equ6}) hold generally for fixed $\alpha$ and do not require $\sigma
_{0}=0 $ or independent payoffs.

Equation (\ref{equ5}) says that the mean of the mixture asset is a weighted average
of $\mu_{0}$ and $\mu_{1}$. Equation (\ref{equ6}) says that the variance does not
have this property; its value is not simply $\alpha\sigma
_{0}^{2}+(1-\alpha)\sigma_{1}^{2}$, but is inflated by an extra term,
$%
\alpha(1-\alpha)(\mu_{1}-\mu_{0})^{2}$, determined by the difference
between the two underlying means.

To satisfy Buridan, the decision-maker must be indifferent between the two
original assets, A and C, and a set of $\alpha$-mixture assets with
parameters $(\sigma_{\alpha},\mu_{\alpha})$ given by (\ref{equ5}) and (\ref{equ6}),
with $%
\alpha$ taking values between 0 and 1. These assets lie on a curve
connecting A and C. The equation of this curve (which is the indifference
curve implied by the Buridan axiom) is found by solving (\ref{equ5}) and (\ref{equ6}) so
as to
eliminate $\alpha$, giving%
%
\begin{equation}
\label{equ7}
\sigma^{2}+\bigl[\mu-(\rho_{0}+\mu_{0})
\bigr]^{2}=\rho_{0}^{2},
\end{equation}
where $\rho_{0}= {\frac12}
[\sigma_{1}^{2}/(\mu_{1}-\mu_{0})+(\mu_{1}-\mu_{0})]$.

Note that the Buridan-based indifference curve (\ref{equ7}) has the form of a circle
in the $(\sigma,\mu)$ plane, with centre at $(0,\mu_{0}+\rho_{0})$ and
radius $\rho_{0}$ (hence the notation). To be sensible, no two
indifference circles can intersect.

Although (\ref{equ7}) represents a full circle, further considerations reveal that
only part of this circle constitutes a sensible indifference curve. The part
of (\ref{equ7}) with $\sigma<0$ may be ignored because negative $\sigma$ does not
exist. The quarter circle DE can also be ignored, because any point on (\ref{equ7})
between D and E is better than D. It has smaller $\sigma$ and larger
$\mu$
and hence must be preferred under the continuity axiom. This leaves as a
plausible indifference curve only the quarter circle AD, of which only the
part between A and C is justified so far, having been derived from the
Buridan axiom.

It remains, therefore, to examine the arc between C and D, for which
$\sigma
_{1}<\sigma\leq\rho_{0}$. Let $\sigma^{\ast}$ fall in this interval,
and consider all assets $(\sigma^{\ast},\mu)$ on the vertical line
through $\sigma=\sigma^{\ast}$. Just as for $\sigma_{1}$, the CMF axiom
requires that there is an asset F on this line with sufficiently high
$\mu$
to make the decision-maker indifferent between F and the risk-free
asset A.
When applied to~$\sigma^{\ast}$, the logic relating to $\sigma_{1}$,
which led to an indifference circle through A and C, produces an
indifference circle through A and F, intersecting the line through BC. This
is possible only when F lies on the original circle through AC. Otherwise
there are two indifference points, both with $\sigma=\sigma_{1}$ yet with
different means~$\mu$. Repeating this argument for all possible
$\sigma
^{\ast}$ in the interval $\sigma_{1}<\sigma^{\ast}\leq\rho_{0}$, the
indifference circle through AC is found to extend to D, thus completing the
quarter circle AD.

Consider next any asset with $\sigma>\rho_{0}$. The decision-maker cannot
be indifferent between a point in this region and A. Buridan's axiom would
have any such point on the same indifference circle as A and, in repeat of
the argument above, there would be contradictions where that curve
intersected any line of constant $\sigma<\rho_{0}$, such as the line
through B and C. More specifically, it follows that all points in the region
$\sigma>\rho_{0}$ must be worse than A. To see this, consider point L
which has the same mean as A yet is worse than A because of its higher
standard deviation. Now imagine that there is some point like G that
has a
mean so large that it is better than A, despite having the same standard
deviation as L. Then by continuity there must be a point between L and G
that is indifferent to A. But again this is impossible because of the
contradictions it would cause with the existing indifference curve. Hence,
all points like L with $\sigma>\rho_{0}$ must be inferior to A, and thus
lie on a lower (larger radius) indifference curve than AD.

Finally, consider assets about the northeast quadrant with respect to D,
for which $\mu>\mu_{0}+\rho_{0}$ and $\sigma>\rho_{0}$. More
particularly, consider three assets labelled H and K and M that define a
rectangle with corners HKMD. Of these, H is preferred to D, since it has
higher mean $\mu^{\prime}>\mu_{0}+\rho_{0}$ and the same standard
deviation $\sigma>\rho_{0}$. Likewise, D is preferred to M because it has
the same mean and lower $\sigma$. Thus, letting $\succ$ symbolize\vadjust{\goodbreak}
``is preferred to'', H${}\succ{}$D${}\succ{}$M.
Similarly, H${}\succ{}$K${}\succ{}$M, by the same reasoning. Unfortunately,
however, these preferences do not complete the rectangle, since they do not
imply any ordering between D and K.

To see the problem here, suppose to begin with \mbox{D${}\succ{}$K}. Then H${}\succ
{}$D${}\succ{}$K, so D is intermediate between K and H. However, K and H are
on a
line of constant mean, so there must be an asset N on this line between H
and K that is indifferent to D, and thus also indifferent to A, thus
contradicting the indifference circle AD already in place. By an identical
argument, this time assuming K${}\succ{}$D (forgetting for a moment that this
ordering has already been shown impossible), there must be some further point
P on the line of constant $\sigma$ between M and K that is indifferent
to D
and A, again contradicting the existing indifference curve AD.

It is impossible, therefore, to resolve all preference relationships within
the rectangle HKMD in a way consistent with Buridan and the CMF axiom. The
only way to avoid this inconsistency is to exclude all assets such as K with
$\mu>\mu_{0}+\rho_{0}$ and $\sigma>\rho_{0}$ from the class of assets
under consideration. In effect, this rules out all points $\mu>\mu
_{0}+\rho
_{0}$ above line RD in Figure~\ref{fig1}, since asset D lies on an indifference
circle centred at $\mu_{0}+\rho_{0}$ and of arbitrary radius. The
implication, therefore, is that it is not possible to rank the class of
\textit{all} possible assets on a MV basis in a way that is consistent with
axioms that would seem essential to any coherent MV decision framework. This
reaffirms the counterexample of \citet{Bor69}, but is reached by a more
general line of reasoning.

\section{Reconciling MV and EU Frameworks}

Contrary to Borch's paradox, it is possible to manufacture sensible
$(\sigma
,\mu)$ indifference curves by either constraining the asset class (in the
way as described above) or by placing other restrictions on the decision
model that limit its theoretical generality and possible practical
relevance. We now discuss the most common ways of forming workable
indifference curves, by which we mean an MV decision framework that yields
the same investment choices (identical rankings of a given set of
distributions) as those based explicitly on~EU.

\subsection{Quadratic Utility}

Mean--variance analysis can be put to work on the generally implausible
assumption of quadratic utility, provided that the returns on the available
assets are\vadjust{\goodbreak} constrained to suit this particular utility function. Since any
utility function has arbitrary location and scale, there is only one
parameter free to select and we may write any risk averse quadratic utility
as $u(x)=2ax-x^{2}$, for some $a>0$, implying $u(0)=0$ and a maximum of
$%
a^{2}$ at $x=a$. Expected utility is then%
%
\begin{eqnarray}
\label{equ8}
\mathit{EU}&=&2a\mu-\bigl(\sigma^{2}+\mu^{2}\bigr)\nonumber\\[-8pt]\\[-8pt]
&=&-
\sigma^{2}-(\mu-a)^{2}+a^{2},\nonumber
\end{eqnarray}
and the indifference curves are circles with centres at $(0,a)$ and various
radii depending on the common fixed EU. Since these circles have been
obtained using expected utility, they automatically satisfy Buridan.

Quadratic utility (QU) exhibits negative marginal utility beyond a
point of
personal satiation. A coherent EU decision framework might nonetheless
assume QU provided that none of the class of assets (distributions) under
consideration offers possible wealth $x$ in the domain where $u(x)$ is
decreasing in $x$. The point of maximum possible quadratic utility, $x=a$,
is represented by $R=(0,a)$ in Figure~\ref{fig2}, where in that instance $a$
equals $%
\mu_{0}+\rho_{0}$.

The condition that no possible payoff $x$ exceeds $a$, where $x$ is
either a
certain or uncertain outcome, implies of course that $\mu\leq a$.
Thus, by
applying this condition concerning possible payoffs $x$ to all admissible
assets, the analysis admits only assets with $\mu<a$, and thus only assets
sitting below asset R (line RD) in Figure~\ref{fig2}. This is the region of the $
(\sigma,\mu)$ half-plane that we found admissible in our axiomatic
critique of MV.

It is important to note, however, that it is not sufficient to exclude any
asset with $\mu\geq a$ [for which $d(\mathit{EU})/d\mu=2(a-\mu)\leq0$] because this
condition is not strong enough to exclude all assets with one or more
\textit{possible} wealth payoffs $x$ in excess of $a$. To avoid any possible
incoherence, the stronger condition of $x\leq a$ for all $x$ must be applied
before the analysis can be conducted in terms of mean and variance.

\begin{figure}

\includegraphics{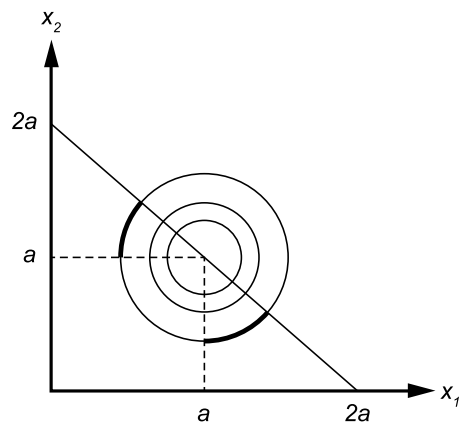}

\caption{Necessary constraint of possible payoffs $x_{1}$ and
$x_{2}$.}\label{fig3}\vspace*{-5pt}
\end{figure}

To see why it is insufficient to exclude assets with $\mu\geq a$, contrary to
conventional shorthand (e.g.,\break Lengwiler, \citeyear{Len04}, page 96), consider a two-point
distribution yielding wealth of either $x=x_{1}$ or $x=x_{2}$ with equal
probability $p=0.5$. The expected quadra\-tic utility is then $%
[(2ax_{1}-x_{1}^{2})+(2ax_{2}-x_{2}^{2})]/2$ and, hence, the
equi-utility $%
(x_{1},x_{2})$ indifference-contours are concentric circles centred at $
(a,a) $, as shown in Figure~\ref{fig3}. Further, with $p=0.5$, the constraint
that $%
\mu< a$ requires that we consider only $(x_{1},x_{2})$ pairs to the left
of the solid diagonal line $(x_{1}+x_{2})/2=a$. The problem is that this
constraint does not remove the $(x_{1},x_{2})$ pairs highlighted by the two
solid thick sections on one of the indifference contours. Clearly, however,
the assets so indicated are not indifferent. In both the dark highlighted
parts of the indifference curve, the rightmost $(x_{1},x_{2})$ pairs are
preferred by any rational decision-maker, because in both cases a shift to
the right means that $x_{1}$ and $x_{2}$ both increase. To avoid this source
of incoherence, it is essential to limit the analysis to asset pairs in the
lower-left quadrant, where neither $x_{1}$ nor $x_{2}$ is greater than $a$.

It is at first disconcerting that this constraint on $x$ did not arise in
our ``first principles'' derivation in Section~\ref{sec6}. The reason for this is that
the axioms on which this analysis is based are too minimal (they
concern $%
\mu$ and $\sigma$ but say nothing directly about $x$) and are insufficient
to reveal the difference between the asset $(x_{1},x_{2})$ pairs
highlighted in Figure~\ref{fig3}. The example depicted in Figure~\ref{fig3} reveals clearly,
however, that it is essential to exclude all such assets before taking
on the
convenience of working in terms of distribution moments $(\sigma,\mu)$.

This example should be seen as an alternative version of Borch's paradox.
Borch relied on his rather opaque counterexample to condemn MV analysis
generally, but the more reasonable conclusion, recognized by Baron, is that
MV analysis can mimic a coherent application of EU under quadratic utility
provided that the asset class is suitably restricted before the
distributional properties of those assets are reduced to their
parameters $%
(\sigma,\mu)$. This is the same restriction as is necessary for coherence
under EU when assuming QU, and reflects a long-known defect\vadjust{\goodbreak} of the quadratic
utility function rather than any flaw in the mathematical restatement of
expected quadratic utility in terms of mean and variance.

It can be argued that the need to exclude assets with potentially
``high''
payoffs from any analysis under quadratic utility (whether via MV or EU
methods) is particularly bothersome, since that is when the decision-maker
may feel most interest in the analysis. This is a problem for QU rather than
for just the MV expression of QU.

\subsection{Normally Distributed Payoffs}

If the class of returns distributions is restricted to a scale-location
family with density $f[(x-k)/b]/b$, given fixed $f$ but variable $k$
and $b$%
, the decision-maker's expected utility will depend only on $k$ and~$b$, and
typically on $\sigma$ and $\mu$. This will lead to indifference
curves in
the $(\sigma,\mu)$-plane. A popular special case is to consider only
assets belonging to the class of normal distributions, $f[(x-\mu
)/\sigma
]/\sigma$. If, for example, we specialize further by considering the
class of
utility functions of constant absolute risk aversion, $u(x)=1-\exp
[-\kappa
x]$, for some $\kappa>0$, the expected utility is easily evaluated to
be%
%
\begin{equation}
\label{equ9}
1-\exp\bigl[-\kappa\mu+\kappa^{2}\sigma^{2}/2\bigr].
\end{equation}
It follows immediately that EU indifference curves in the $(\sigma
^{2},\mu
) $-plane are straight lines, $\mu-\kappa\sigma^{2}/2=\mathrm{constant}$ (where
the higher the constant, the higher the EU). In the $(\sigma,\mu)$ plane,
these same curves appear as parabolas, all with axes $\sigma=0$ and
increasing in $\mu$ as $\sigma$ increases. These contain none of the
inherent contradictions revealed in the case of the circular indifference
curves derived from either Buridan or the assumption of quadratic utility.

It is important to understand how we have arrived at a case of parabolic
indifference curves in apparent contradiction with the circles based on
Buridan. These different families of indifference curves emanate from
different starting points. To get the Buridan (circular) curves, we presume
what is implicit under quadratic utility---specifically, that all available
assets (``mixed'' or ``pure'') are represented sufficiently by just $(\sigma,\mu
) $. Similarly, to arrive at the normal-constant-risk-aversion (parabolic)
indifference curves, we make a contrary and more restrictive assumption,
namely, that all possible assets have normal distributions.

Derivation of indifference curves on the assumption of strictly normal
distributions does not sit well with Buridan's axiom. This class of
distributions is not closed under probability\vadjust{\goodbreak} mixing, since probability
mixtures of normals are typically not normal. It is self-defeating,
therefore, to arbitrarily limit the admissible asset class to just normal
distributions, since mixture assets can always be constructed by
randomization whether or not they arise naturally.

From a utility theory standpoint, this issue can be summed up as
follows. If
normal assets with parameters $(\sigma_{1},\mu_{1})$ and $(\sigma
_{2},\mu
_{2})$ have expected utilities $u_{1}=V(\sigma_{1},\mu_{1})$ and $%
u_{2}=V(\sigma_{2},\mu_{2})$, then any probability mixture thereof,
defined by $\alpha$, has expected utility $u_{\alpha}=\alpha
u_{1}+(1-\alpha)u_{2}$. Such calculations are elementary to expected
utility theory, yet cannot be captured or exhibited in any way using a set
of indifference curves applicable to only normal distributions. The
parameters of a nonnormal probability mixture, $(\sigma_{\alpha},\mu
_{\alpha})$, are known but meaningless, since they cannot be substituted
into (\ref{equ9}) or any other measure of the expected utility of a normal
distribution. They can, of course, be substituted into (\ref{equ8}), but that would
presume quadratic utility rather than normality.

\subsubsection{Chipman--Baron defence of normality}

Having first revealed that Borch's paradox exploits a defect in QU rather
than one in MV per se, \citet{Bar77} went on to rationalize the use of MV
indifference curves under the common presumption of only normal
distributions. This further contribution, elaborating upon Chipman [(\citeyear{Chi73}),
pages 179--181] rests primarily on the assumption that assets are all highly
divisible, thereby allowing investors to hold them in conventional linearly
weighted portfolios with arbitrary positive weights (e.g., the
investor can buy \$3.121 worth of asset A and \$6.879 worth of asset B
in a
\$10 portfolio).

Linearly weighted portfolios have two helpful properties. First, as is
well known, the class of jointly normal distributions is closed under
linear combination. Second, and essential to the Chipman--Baron
argument, it follows from Jensen's inequality that for any increasing
and strictly concave (risk averse) utility function $u(x)$, the
expected utility of an $\alpha$-mixture of any two assets A and B is
less than the expected utility of the corresponding $\alpha$-weighted
portfolio of the same two assets, $\alpha \in(0,1)$. Thus, even for
risk averse utility functions that cannot be represented in terms of
solely $(\sigma,\mu)$, there is always a conventional weighted
portfolio of A and B that dominates any probability mixture of A and B.
Probability mixtures can thus be ignored by any rational risk averse
expected utility maximizer, provided of course that it is possible to
combine the available pure assets into arbitrarily weighted portfolios.\vadjust{\goodbreak}

By excluding all probability mixture assets, Chipman and Baron got around
the problem that mixtures of normals are not normal, and thus made it
feasible to work on the assumption that the asset class worthy of
consideration includes only jointly normal distributions. This leaves the
problem, of course, of how to choose between such normally distributed assets
(both the pure assets and their\break weighted portfolios).

On this issue, Baron emphasized two points. First, the assumption that all
assets are normal allows deci\-sion-makers with essentially any increasing
utility function to draw $(\sigma,\mu)$ indifference curves. For instance,
a decision-maker with exponential utility has parabolic $(\sigma,\mu)$
indifference curves. Second, it is wrong to draw an arbitrary set of
convex $%
(\sigma,\mu)$ indifference curves and presume that they derive from some
sensible risk averse utility function. To the contrary, as occurs in Borch's
paradox, a set of apparently quite plausible looking indifference curves
drawn on the basis of intuition, rather than being derived from an
underlying utility function $u(x)$, will generally embody preferences
between at least some definable assets that are obviously irrational.

This second point, due initially to Chipman\break [(\citeyear{Chi73}), pages
168 and 169] is not widely known. It is commonplace, especially in
classroom and textbook contexts, to draw up any ``sensible looking''
usually convex $(\sigma,\mu)$ indifference curves, as if the investor
has free reign to take on any curves she likes. This fundamental
misconception underlines why time can be well spent revisiting the
analytical literature on MV versus EU from the Borch era. The depth of
this literature is exhibited in the way that Chipman
[(\citeyear{Chi73}), page 169] was able to characterize the MV
indifference curves from which an expected utility function can be
recovered. Specifically, Chipman proved that when the choice is between
only normal distributions, the utility function $u(x)$ must be bounded
over $-\infty<x<\infty$ by the condition that $\llvert u(x)\rrvert \leq
a\exp(bx^{2})$, $a,b>0$ so as to ensure that the expected utility
integral converges. Given this growth constraint on $u(x)$, there
exists an indifference function $V(\sigma,\mu)=E[u(x)]$ under the
necessary and sufficient condition that $V(\sigma,\mu)$ satisfies the
differential equation%
\[
\frac{1}{\sigma}\frac{\partial V}{\partial\sigma}=\frac{\partial
^{2}V}{%
\partial\mu^{2}}.
\]

\subsubsection{Mixture assets in practice}

Baron [(\citeyear{Bar77}), pa\-ge~1692] and Liu [(\citeyear{Liu04}), page
233] discuss how probability mixtures of different assets occur,
explicitly or implicitly. For instance,\vadjust{\goodbreak} the cash flows from a firm
might have different probability distributions depending on a random
event such as the outcome of a law suit, the reaction of competitors or
a regulatory or political shift.

There is no stock market for explicit mixtures of different individual
stocks. Indeed, Baron's argument suggests that there could be little if any
rational demand for these. As an interesting aside, however, in some betting
markets there is a commercially successful product called a ``mystery bet'',
where the buyer agrees to be allocated a random bet of agreed amount
(e.g., a
\$10 bet on a random horse in a random race).

A fully subjectivist view of real world stock market investment would
suggest that much ``rational investment'' is tantamount to making ``mystery
bets'', in that there are so many random factors outside the control or
observation of the investor that determine how the payoffs from her chosen
investments are distributed. On this view, every discrete asset (company
stock) in the stockmarket is in fact a mixture of distributions, and the
investor has a subjective assessment of that stock's payoffs which amounts
to a subjective mixture distribution of latent ``underlying stocks''.

\section{Borch and Stochastic Dominance}

The axioms of EU do not insist that the decision-maker should prefer more
rather than less. To make this basic presumption of human behavior, the
utility function $u(x)$ must be a monotonically increasing function of the
payoff $x$. Once this assumption is made of $u(x)$, it is possible to argue
either on the grounds of utility theory or mere common sense that some
distributions of payoffs strictly dominate others. To begin with, if one
distribution $f_{1}(x)$ is entirely to the right of another $f_{2}(x)$, then
any investor who prefers more to less will favour $f_{1}(x)$. A~little less
obviously, the same order of strict preference holds whenever the cumulative
probability of any given outcome $x\geq x^{\ast}$ is higher under
$f_{1}(x)$ than under $f_{2}(x)$ for all $x^{\ast}$. Specifically, $f_{1}$
is weakly preferred to $f_{2}$ whenever $F_{2}(x^{\ast})-F_{1}(x^{\ast
})\geq0$ for all $x^{\ast}$, where $F(x)$ is the cumulative
distribution function corresponding to probability density $f(x)$.

This condition is known following \citet{HadRus69} as first order
stochastic dominance (FSD). Its implication is that the expected
utility of
distribution $f_{1}(x)$ exceeds the expected utility of distribution $%
f_{2}(x)$ for all strictly increasing $u(x)$. \citet{HadRus69} also
proved that if the class of decision-makers\vadjust{\goodbreak} is limited to only those with
risk averse $u(x)$, $f_{1}$ is weakly preferred to $f_{2}$ if and only
if $%
\int_{-\infty}^{a} [ F_{2}(x)-F_{1}(x) ]
\,dx\geq0$
for all $a>-\infty$ (the potential payoffs are $-\infty\leq
x\leq\infty$). This is known as second order stochastic dominance
(SSD).

The conditions of stochastic dominance discovered by
\citet{HadRus69}, and independently by \citet{HanLev69},
proceed by constraining not the class of asset distributions $f(x)$,
but rather the class of decision-makers. Decision-makers are
categorized in effect by the Pratt--Arrow measure of absolute risk
aversion $r(x)=-u^{\prime\prime }(x)/u^{\prime}(x)$, which is constant
under positive linear transformations of $u(x)$ and hence characterizes
decision-makers uniquely [i.e., such that any two decision-makers with
the same preferences have the same $r(x)$]. FSD is an implicit
dominance criterion for any decision-maker with increasing twice
differentiable utility\break $(-\infty<r(x)<\infty )$, whereas SSD is
implicit in the decision criteria of any risk averse decision-maker
$(0<r(x)<\infty )$. \citet{Mey77} provides a succinct overview and
generalization of the \citet{HadRus69} and \citet{HanLev69}
proofs. See Levy (\citeyear{Lev06}) for a comprehensive synthesis of the theory of
stochastic dominance as a framework for decision-making under
uncertainty. Levy [(\citeyear{Lev12}),\break Chapter 3] shows that with normal
distributions, SSD is equivalent to the MV criterion, and can be used
as the axiomatic basis on which to derive it.

\citet{Fis80} made significant strides towards a general theory of ranking
assets in order of stochastic dominance using only their moments. He proved,
for example, that (i) first order stochastic dominance of asset A over asset
B implies that $\mu_{A}>\mu_{B}$, and (ii) second order stochastic
dominance of asset A over asset B implies that $\mu_{A}\geq\mu_{B}$
and $\sigma_{A}<\sigma_{B}$. Similar conditions involving higher moments
follow necessarily under higher orders of stochastic dominance. These proofs
have not led to a theory of ranking all assets via moments because the
moment conditions are necessary but not sufficient to prove stochastic
dominance at their respective levels. This has recently been emphasized by
Liu [(\citeyear{Liu04}),\break pages~231 and 232] who gave a general proof that there is no specifiable
set of moment conditions concerning the first $n$ moments of A and B that
imply a first, second or third order dominance relationship between those
two assets. Essentially, while certain moment conditions are suggestive of
certain orders of stochastic dominance, those conditions can still arise
when their corresponding order of dominance does not obtain.

\citet{LevSar69} employed the principle of FSD to rebut Borch. They
explained that the basic mistake in Borch's logic is to treat the joint
conditions that $\mu_{1}>\mu_{2}$ and $\sigma_{1}>\sigma_{2}$ as if
these alone are sufficient for two assets $(\sigma_{1},\mu_{1})$ and $
(\sigma_{2},\mu_{2})$ to be seen by someone as indifferent and positioned
on the same indifference curve. Their counterargument is that while
$\mu
_{1}>\mu_{2}$ and $\sigma_{1}>\sigma_{2}$ are necessary under risk
aversion for indifference, these conditions are not sufficient. So much is
proven by Borch's own example, in which two specified assets meeting both
conditions are obviously not indifferent.

\citet{LevSar69} go on to reveal that one of the two assets in Borch's
example is FSD over the other. Again, this is clear since one asset
produces $%
x$ or $y_{1}$ with probability $p$ and the other produces $x$ or $y_{2}$
with the same probability, implying that if $y_{1}>y_{2}$, for instance,
then the cumulative distribution of asset~1 is weakly to the right of asset
2. The Levy and Sarnat antidote to Borch is thus to adopt a modified
application of MV where in the first step the admissible asset class is
immediately and efficiently reduced by removing all assets that are
dominated under FSD (and hence are not in the efficient set for \textit{any}
decision-maker with increasing utility). This two-stage procedure was first
proposed in \citet{HanLev69} and later by \citet{Lev74} and Levy and
Sarnat (\citeyear{LevSar72}), pages 315--318. There can be no argument against FSD from any
angle. In the words of a referee, it is ``normatively desirable and
descriptively accurate''.

\section{The Capital Asset Pricing Model}

The restatement of a subset of EU in the language of MV led immediately to
formulation of the capital asset pricing model (CAPM) by \citet{Sha64},
\citet{Lin65} and Mossin (\citeyear{Mos66}). In this section we set out a simple
derivation of the CAPM and explain briefly what it offers that might not
have been discovered without MV.

There are $n$ risky assets in the market and the price of asset $j$ is $
P_{j} $ $(j=1,2,\ldots,n)$. The investor spreads her money between risky assets
and risk-free bonds. Her portfolio weights are therefore $w_{\mathrm{rf}}$ in the
risk-free asset and $w_{M}=1-w_{\mathrm{rf}}$ in the market. Her investment
in the
market (risky assets) is spread evenly across all $n$ risky assets in
proportion to their respective prices. She earns return $r_{\mathrm{rf}}$ from the
risk free asset and $r_{M}$ from the market portfolio of risky assets. By
definition, the return on the market portfolio is $r_{M}=\sum_{j}P_{j}r_{j}/\sum_{j}P_{j}$, where $r_{j}$ is the
return on asset $j$.\vadjust{\goodbreak}

Suppose that $r_{\mathrm{rf}}$ is less than both expected returns $\mu(r_{M})$
and $%
\mu(r_{j})$, as must be the case to attract risk averse investors. To
increase the expected return of her investment portfolio, the first
possibility is that the investor buys some more of asset $j$ using some of
the money she has invested in the risk-free asset. Her new portfolio weights
are then $w_{M}$ in the market portfolio, $\delta$ in security $j$ and
$%
w_{\mathrm{rf}}-\delta$ in the risk-free asset. The expected return of this
portfolio is $w_{M}\mu(r_{M})+\delta\mu(r_{j})+(w_{\mathrm{rf}}-\delta)r_{\mathrm{rf}}$
and its variance is $w_{M}^{2}\sigma^{2}(r_{M})+\delta^{2}\sigma
^{2}(r_{j})+2w_{M}\delta\operatorname{cov}(r_{j},r_{M})$. The marginal increase in
expected return is therefore $\delta\mu(r_{j})-\delta r_{\mathrm{rf}}$. Similarly,
the marginal increase in portfolio variance is $\delta^{2}\sigma
^{2}(r_{j})+2w_{M}\delta\operatorname{cov}(r_{j},r_{M})$, which approaches
$2w_{M}\delta
\operatorname{cov}(r_{j},r_{M})$ for small $\delta$. The marginal rate of substitution,
or price in terms of added risk (variance) for each extra unit of expected
return (mean), is thus%
%
\begin{equation}
\label{equ10}
\frac{\mu(r_{j})-r_{\mathrm{rf}}}{2w_{M}\operatorname{cov}(r_{j},r_{M})}.
\end{equation}
The second way for the investor to increase expected return is to sell
weight $\delta$ of the risk-free asset and add weight $\delta$ to her
investment in the market portfolio. By an identical argument to that above,
the marginal rate of substitution is then%
%
\begin{equation}
\label{equ11}
\frac{\mu(r_{M})-r_{\mathrm{rf}}}{2w_{M}\sigma^{2}(r_{M})}.
\end{equation}

Setting $\mbox{(\ref{equ10})}=\mbox{(\ref{equ11})}$, on the basis that there cannot be two different unit
prices to achieve the same result in a rational market (the ``law of no
arbitrage''), leads to the equation commonly known as the mean--variance
CAPM%
%
\begin{equation}
\label{equ12}
\mu(r_{j})=r_{\mathrm{rf}}+\frac{\operatorname{cov}(r_{j},r_{M})}{\sigma
^{2}(r_{M})} \bigl[
\mu(r_{M})-r_{\mathrm{rf}} \bigr].
\end{equation}

To rewrite this equation in terms of asset prices rather than returns,
let the return on asset $j$ be defined in terms of its initial price
$P_{i}$ and its period-end price or value $V_{j}$ by
$r_{j}=V_{j}/P_{j}-1$. Hence, by definition,
$\mu(r_{i})=\mu(V_{i})/P_{i}-1$,
$\operatorname{cov}(r_{i},r_{M})=\break\operatorname{cov}
(V_{j},V_{M})/P_{j}P_{M}$ and $\sigma(r_{M})=\sigma(V_{M})/P_{M}$.
Substituting in (\ref{equ12}) and rearranging reveals the CAPM as an
explicit pricing
model%
\begin{eqnarray}
P_{j}=\frac{\mu(V_{j})-\beta_{j} [ \mu
(V_{M})-P_{M}(1+r_{\mathrm{rf}}) ]
}{1+r_{\mathrm{rf}}}\nonumber\\
&&\eqntext{\displaystyle \biggl(P_{M}=\sum
_{j}P_{j}\biggr),}
\end{eqnarray}
where $\beta_{j}=\operatorname{cov}(V_{j},V_{M})/\sigma^{2}(V_{M})$.\vadjust{\goodbreak}

The CAPM asset prices $P_{j}$ can be understood as either (i) coherent
prices in the mind of a given investor with quadratic utility, or (ii)
market equilibrium prices in a market where all investors have quadratic
utility and the same probability beliefs regarding the uncertain future
asset values $V_{j}$. The wider possibility drawing on the Chipman--Baron
argument is that only normally distributed assets need be considered, in
which case the investor(s) need not have quadratic utility.

The appeal of the CAPM equation is that rather than looking at assets
one by
one, each risky asset is valued with respect to what it contributes to an
optimally weighted portfolio. Upon taking this unified rather than piecemeal
approach, the CAPM reveals the factors, most interestingly $\beta_{j}$,
that make each asset more or less valuable to decision-makers seeking an
optimal mix of investments. Particularly revealing is that the risk of
investing in a given asset is measured not by its variance, but by its
covariance with all other risky assets. An asset can therefore be highly
risky of itself and yet still be highly attractive. Furthermore, even
an asset
with negative expected return might have a high price $P_{j}$ if its returns
have negative correlation with the market.

\citet{Bor79} was of course well aware of the CAPM, which he described as
having in finance something like the status of $E=mc^{2}$. He was critical
of the CAPM for the fact that it stands on MV foundations and is therefore
open to the same kind of ``nonsense results'' as generalized MV
decision-making. Anticipating what has more recently become very widely
accepted,
\citet{Bor79} noted that the CAPM did not perform well when fitted to actual
market price data. \citet{Lev12} has recently reviewed the history and status
of the CAPM, with emphasis on how it has survived as a highly important tool
in financial practice while at the same time having its empirical or
descriptive validity widely disparaged.

A more sympathetic view of the CAPM (\cite{Mey87}, page 426) is that by
effectively restating expected quadratic utility in terms of moments $%
(\sigma,\mu)$, finance theorists uncovered a coherence relationship
between asset prices that had not been evident from the higher plane of EU
theory. Reassuringly, much of the ``new finance'' inspired by MV and the
CAPM hinges on the rule of
no arbitrage or ``no Dutch book''. This
idea appeared much earlier in the writings of de Finetti. Indeed, it would
come as no great surprise to find that de Finetti envisaged a
``subjectivist CAPM''. In principle, a de
Finetti style CAPM would connect the utility\vadjust{\goodbreak} functions and subjective
probability distributions of all investors to a theoretical (and possibly
observable) set of equilibrium asset prices, and would do much to unify
finance and decision analysis at a philosophical level.

\section{Conclusion}

Mean--variance is the most influential theory in the practice of
investment analysis and business deci\-sion-making. This is remarkable
given that Marko\-witz's model is effectively a diminished form of
expected utility theory. Of itself, in the elegant structures set out
by \citet{vonMor53}, \citet{Sav54} and other decision
theorists, utility theory is hardly known to financial practitioners.
This is despite its intellectual traditions and theoretical acceptance
in many fields, including somewhat paradoxically economics.

From the 1950s when Markowitz introduced what he saw as a new theory of
decision-making under subjective probability, there has been a concerted
\mbox{intellectual} undertaking in financial economics towards understanding how
mean--variance methods might be reconciled with expected utility. The
literature on this topic is extremely wide and the task of surveying its
current state and its links to modern financial practice is well beyond what
we have attempted in this paper. An historically thorough and helpful survey
of the literature exists in \citet{Liu04}.

We focus on the historical debate concerning mean--variance and
expected utility, particularly on the once prominent \textit{reductio
absurdum} of mean--variance known as Borch's paradox. This logical
counter-argu\-ment directed by Borch at mean--variance has been rebutted,
most explicitly by \citet{Bar77}. Despite its hidden weak spots,
Borch's paper, and the literature that it initiated, holds timeless
philosophical interest. Opinions are widely divided on whether expected
utility is ``too normative'' to be practical or mean--variance is too
``practical'' to be respectable. The pragmatists' perspective is summed
up by economics Nobel Laureate James Tobin [(\citeyear{Tob69}), page 14]
who suggested that a business practitioner will not be amused by the
instruction that ``he should consult his utility and his subjective
probabilities and then maximize''. Yet contrary of Tobin's mockery,
there is intensive theoretical and empirical study in current finance
research devoted \mbox{precisely} to practical investment portfolio selection
by optimization of certain expected utility functions, both directly
and by their expression through higher moments (e.g.,
\cite{MacZieLi05}; \cite{CreKriPag05}; \cite{Sha07}; \cite{AdlKri07};
\cite{Hagetal08}).

A conciliatory note among the seminal contributors to the MV versus EU
literature was struck by Meyer (\citeyear{Mey87}), page 426. He saw that MV offered
a way
of rewriting a subclass of expected utility that not only simplified the
notions of risk and return, but which also revealed previously unrecognized
relationships between the risks and returns of individual assets and their
combinations in weighted portfolios. This gave rise to the new language of
the ``efficient frontier'' and ultimately the ``capital asset pricing model''.
More fundamentally, it revealed very interesting and sometimes
counterintuitive financial principles. In many instances these results can
be transported back into EU theory and generalized to suit different
possible utility functions. See, for example, the log utility CAPM
applied in
\citet{Joh12}. This asset pricing equation is obtainable from first
principles for any $E[\log(x)]$ maximizing portfolio optimizer. Such
equations might have been discovered without being triggered by Markowitz
and the invention of the mean--variance CAPM, but more realistically the
academic field of asset pricing was inspired by Markowitz and the theory
that arose in the era of Borch and the swinging 60s.

\section{Postscript}

After completing this survey of Borch's paradox, the authors learnt
more of
the status that Borch retains in actuarial science from a fascinating
biography by \citet{Aas03}. The following passage from this biography confirms
much of the impression we gained from the literature regarding Borch
and his
intellectual influence over the mean--variance debate:

\begin{quotation}
There is a story about Borch's stand on ``mean--variance'' analysis.
This story is known to econ\-omists, but probably unknown to actuaries:
He published a paper, ``A note on Uncertainty and
Indifference Curves'' in Review of Economic \mbox{Studies}
(1969), and Martin Feldstein, a~friend of Borch, published another
paper in the same issue on the limitations of the mean--vari\-ance
analysis for portfolio choice (Feldstein, \citeyear{Fel69}). In the same issue a
comment from James Tobin appeared, ``Comment on Borch
and Feldstein'' (\cite{Tob69}). Today Borch's and
Feldstein's criticism seems well in place, but at the time this was
shocking news. In particular, Professor James\vadjust{\goodbreak} Tobin at Yale, later a
Nobel laureate in economics, entertained at the time great plans for
incorporating mean--variance analysis in macroeconomic modelling. There
was even financing in place for an institute on a national level.
However, after Borch's and Feldstein's papers were published, Tobin's
project seemed to have been abandoned. After this episode, involving
two of the leading American economists, Borch was well noticed by the
economist community, and got a reputation, perhaps an unjust one, as a
feared opponent.
\end{quotation}

\section*{Acknowledgement}

The authors thank Jay Kadane and anonymous referees for helpful comments.




\begin{thebibliography}{56}

\bibitem[\protect\citeauthoryear{Aase}{2004}]{Aas03}
\begin{bincollection}[auto:STB|2013/03/04|13:35:07]
\bauthor{\bsnm{Aase},~\bfnm{K.~K.}\binits{K.~K.}}
(\byear{2004}).
\btitle{The life and career of Karl H. Borch}.
In \bbooktitle{Encyclopedia of Actuarial Science. Vol. 1}
(\beditor{\bfnm{J.~L.}\binits{J.~L.}~\bsnm{Teugels}} \AND
  \beditor{\bfnm{B.}\binits{B.}~\bsnm{Sundt}}, eds.)
\bpages{191--195}.
\bpublisher{Wiley}, \blocation{New York}.
\bptok{imsref}%
\end{bincollection}
\endbibitem

\bibitem[\protect\citeauthoryear{Adler and Kritzman}{2007}]{AdlKri07}
\begin{barticle}[auto:STB|2013/03/04|13:35:07]
\bauthor{\bsnm{Adler},~\bfnm{T.}\binits{T.}} \AND
  \bauthor{\bsnm{Kritzman},~\bfnm{M.}\binits{M.}}
(\byear{2007}).
\btitle{Mean--variance versus full-scale optimization: In and out of sample}.
\bjournal{Journal of Asset Management}
\bvolume{7}
\bpages{302--311}.
\bptok{imsref}%
\end{barticle}
\endbibitem

\bibitem[\protect\citeauthoryear{Baron}{1977}]{Bar77}
\begin{barticle}[auto:STB|2013/03/04|13:35:07]
\bauthor{\bsnm{Baron},~\bfnm{D.~P.}\binits{D.~P.}}
(\byear{1977}).
\btitle{On the utility theoretic foundations of mean--variance analysis}.
\bjournal{J. Finance}
\bvolume{32}
\bpages{1683--1697}.
\bptok{imsref}%
\end{barticle}
\endbibitem

\bibitem[\protect\citeauthoryear{Barone}{2008}]{Bar08}
\begin{barticle}[mr]
\bauthor{\bsnm{Barone},~\bfnm{Luca}\binits{L.}}
(\byear{2008}).
\btitle{Bruno de {F}inetti and the case of the critical line's last segment}.
\bjournal{Insurance Math. Econom.}
\bvolume{42}
\bpages{359--377}.
\bid{doi={10.1016/j.insmatheco.2007.04.003}, issn={0167-6687}, mr={2392094}}
\bptok{imsref}%
\end{barticle}
\endbibitem

\bibitem[\protect\citeauthoryear{Barucci}{2003}]{Bar03}
\begin{bbook}[mr]
\bauthor{\bsnm{Barucci},~\bfnm{Emilio}\binits{E.}}
(\byear{2003}).
\btitle{Financial Markets Theory:
Equilibrium, Efficiency and Information}.
\bpublisher{Springer}, \blocation{London}.
\bid{doi={10.1007/978-1-4471-0089-8}, mr={1958149}}
\bptok{imsref}%
\end{bbook}
\endbibitem

\bibitem[\protect\citeauthoryear{Bernardo and Smith}{1994}]{BerSmi94}
\begin{bbook}[mr]
\bauthor{\bsnm{Bernardo},~\bfnm{Jose-M.}\binits{J.-M.}} \AND
  \bauthor{\bsnm{Smith},~\bfnm{Adrian F.~M.}\binits{A.~F.~M.}}
(\byear{1994}).
\btitle{Bayesian Theory}.
\bpublisher{Wiley}, \blocation{Chichester}.
\bid{doi={10.1002/9780470316870}, mr={1274699}}
\bptok{imsref}%
\end{bbook}
\endbibitem


\bibitem[\protect\citeauthoryear{Borch}{1969}]{Bor69}
\begin{barticle}[auto:STB|2013/03/04|13:35:07]
\bauthor{\bsnm{Borch},~\bfnm{K.}\binits{K.}}
(\byear{1969}).
\btitle{A note on uncertainty and indifference curves}.
\bjournal{Rev. Econom. Stud.}
\bvolume{36}
\bpages{1--4}.
\bptok{imsref}%
\end{barticle}
\endbibitem

\bibitem[\protect\citeauthoryear{Borch}{1973}]{Bor73}
\begin{barticle}[auto:STB|2013/03/04|13:35:07]
\bauthor{\bsnm{Borch},~\bfnm{K.}\binits{K.}}
(\byear{1973}).
\btitle{Expected utility expressed in terms of moments}.
\bjournal{Omega: The International Journal of Management Science}
\bvolume{1}
\bpages{331--343}.
\bptok{imsref}%
\end{barticle}
\endbibitem

\bibitem[\protect\citeauthoryear{Borch}{1974}]{Bor74}
\begin{barticle}[auto:STB|2013/03/04|13:35:07]
\bauthor{\bsnm{Borch},~\bfnm{K.}\binits{K.}}
(\byear{1974}).
\btitle{The rationale of the mean--standard deviation analysis: Comment}.
\bjournal{American Economic Review}
\bvolume{64}
\bpages{428--430}.
\bptok{imsref}%
\end{barticle}
\endbibitem

\bibitem[\protect\citeauthoryear{Borch}{1978}]{Bor78}
\begin{barticle}[auto:STB|2013/03/04|13:35:07]
\bauthor{\bsnm{Borch},~\bfnm{K.}\binits{K.}}
(\byear{1978}).
\btitle{Portfolio theory is for risk lovers}.
\bjournal{Journal of Banking and Finance}
\bvolume{2}
\bpages{179--181}.
\bptok{imsref}%
\end{barticle}
\endbibitem

\bibitem[\protect\citeauthoryear{Borch}{1979}]{Bor79}
\begin{barticle}[mr]
\bauthor{\bsnm{Borch},~\bfnm{Karl}\binits{K.}}
(\byear{1979}).
\btitle{Equilibrium in capital markets}.
\bjournal{Econom. Lett.}
\bvolume{2}
\bpages{175--179}.
\bid{doi={10.1016/0165-1765(79)90169-1}, issn={0165-1765}, mr={0539182}}
\bptok{imsref}%
\end{barticle}
\endbibitem

\bibitem[\protect\citeauthoryear{Chipman}{1973}]{Chi73}
\begin{barticle}[auto:STB|2013/04/24|11:25:54]
\bauthor{\bsnm{Chipman},~\bfnm{J.~S.}\binits{J.~S.}}
(\byear{1973}).
\btitle{The ordering of portfolios in terms of mean and variance}.
\bjournal{Rev. Econom. Stud.}
\bvolume{40}
\bpages{167--190}.
\bptok{imsref}%
\end{barticle}
\endbibitem

\bibitem[\protect\citeauthoryear{Cochrane}{2001}]{Coc01}
\begin{bbook}[auto:STB|2013/03/04|13:35:07]
\bauthor{\bsnm{Cochrane},~\bfnm{J.}\binits{J.}}
(\byear{2001}).
\btitle{Asset Pricing}.
\bpublisher{Princeton Univ. Press}, \blocation{Princeton}.
\bptok{imsref}%
\end{bbook}
\endbibitem

\bibitem[\protect\citeauthoryear{Cremers, Kritzman and
  Page}{2005}]{CreKriPag05}
\begin{barticle}[auto:STB|2013/03/04|13:35:07]
\bauthor{\bsnm{Cremers},~\bfnm{J.~H.}\binits{J.~H.}},
  \bauthor{\bsnm{Kritzman},~\bfnm{M.}\binits{M.}} \AND
  \bauthor{\bsnm{Page},~\bfnm{S.}\binits{S.}}
(\byear{2005}).
\btitle{Optimal hedge fund allocations}.
\bjournal{Journal of Portfolio Management}
\bvolume{31}
\bpages{70--81}.
\bptok{imsref}%
\end{barticle}
\endbibitem

\bibitem[\protect\citeauthoryear{de~Finetti}{1940}]{deF40}
\begin{barticle}[mr]
\bauthor{\bparticle{de} \bsnm{Finetti},~\bfnm{B.}\binits{B.}}
(\byear{1940}).
\btitle{Il problema dei ``{P}ieni.''}
\bjournal{Giorn. Ist. Ital. Attuari}
\bvolume{11}
\bpages{1--88}.
\bnote{English Transaltion by Luca Barone in \textit{Journal of
Investment Management} \textbf{4} (2006) 19--43}.
\bptok{imsref}%
\end{barticle}
\endbibitem

\bibitem[\protect\citeauthoryear{DeGroot}{1970}]{DeG70}
\begin{bbook}[mr]
\bauthor{\bsnm{DeGroot},~\bfnm{Morris~H.}\binits{M.~H.}}
(\byear{1970}).
\btitle{Optimal Statistical Decisions}.
\bpublisher{McGraw-Hill}, \blocation{New York}.
\bid{mr={0356303}}
\bptok{imsref}%
\end{bbook}
\endbibitem

\bibitem[\protect\citeauthoryear{Eeckhoudt, Gollier and
  Schlesinger}{2005}]{EecGolSch05}
\begin{bbook}[auto:STB|2013/03/04|13:35:07]
\bauthor{\bsnm{Eeckhoudt},~\bfnm{L.}\binits{L.}},
  \bauthor{\bsnm{Gollier},~\bfnm{C.}\binits{C.}} \AND
  \bauthor{\bsnm{Schlesinger},~\bfnm{H.}\binits{H.}}
(\byear{2005}).
\btitle{Economic and Financial Decisions Under Risk}.
\bpublisher{Princeton Univ. Press}, \blocation{Princeton}.
\bptok{imsref}%
\end{bbook}
\endbibitem

\bibitem[\protect\citeauthoryear{Feldstein}{1969}]{Fel69}
\begin{barticle}[auto:STB|2013/03/04|13:35:07]
\bauthor{\bsnm{Feldstein},~\bfnm{M.~S.}\binits{M.~S.}}
(\byear{1969}).
\btitle{Mean--variance analysis in the theory of liquidity preference and
  portfolio selection}.
\bjournal{Rev. Econom. Stud.}
\bvolume{36}
\bpages{5--12}.
\bptok{imsref}%
\end{barticle}
\endbibitem

\bibitem[\protect\citeauthoryear{Fishburn}{1980}]{Fis80}
\begin{barticle}[mr]
\bauthor{\bsnm{Fishburn},~\bfnm{Peter~C.}\binits{P.~C.}}
(\byear{1980}).
\btitle{Stochastic dominance and moments of distributions}.
\bjournal{Math. Oper. Res.}
\bvolume{5}
\bpages{94--100}.
\bid{doi={10.1287/moor.5.1.94}, issn={0364-765X}, mr={0561157}}
\bptok{imsref}%
\end{barticle}
\endbibitem

\bibitem[\protect\citeauthoryear{Hadar and Russell}{1969}]{HadRus69}
\begin{barticle}[auto:STB|2013/03/04|13:35:07]
\bauthor{\bsnm{Hadar},~\bfnm{J.}\binits{J.}} \AND
  \bauthor{\bsnm{Russell},~\bfnm{W.~R.}\binits{W.~R.}}
(\byear{1969}).
\btitle{Rules for ordering uncertain prospects}.
\bjournal{American Economic Review}
\bvolume{59}
\bpages{25--\break34}.
\bptok{imsref}%
\end{barticle}
\endbibitem

\bibitem[\protect\citeauthoryear{Hagstromer et~al.}{2008}]{Hagetal08}
\begin{barticle}[auto:STB|2013/03/04|13:35:07]
\bauthor{\bsnm{Hagstromer},~\bfnm{B.}\binits{B.}},
  \bauthor{\bsnm{Anderson},~\bfnm{R.~G.}\binits{R.~G.}},
  \bauthor{\bsnm{Binner},~\bfnm{J.~M.}\binits{J.~M.}},
  \bauthor{\bsnm{Elger},~\bfnm{T.}\binits{T.}} \AND
  \bauthor{\bsnm{Nilsson},~\bfnm{B.}\binits{B.}}
(\byear{2008}).
\btitle{Mean--variance versus full-scale optimization: Broad evidence for the
  UK}.
\bjournal{The Manchester School}
\bvolume{76}
\bpages{(Supplement) 134--156}.
\bptok{imsref}%
\end{barticle}
\endbibitem

\bibitem[\protect\citeauthoryear{Hanoch and Levy}{1969}]{HanLev69}
\begin{barticle}[auto:STB|2013/03/04|13:35:07]
\bauthor{\bsnm{Hanoch},~\bfnm{G.}\binits{G.}} \AND
  \bauthor{\bsnm{Levy},~\bfnm{H.}\binits{H.}}
(\byear{1969}).
\btitle{The efficiency analysis of choices involving risk}.
\bjournal{Rev. Econom. Stud.}
\bvolume{36}
\bpages{335--346}.
\bptok{imsref}%
\end{barticle}
\endbibitem

\bibitem[\protect\citeauthoryear{Hanoch and Levy}{1970}]{HanLev70}
\begin{barticle}[auto:STB|2013/03/04|13:35:07]
\bauthor{\bsnm{Hanoch},~\bfnm{G.}\binits{G.}} \AND
  \bauthor{\bsnm{Levy},~\bfnm{H.}\binits{H.}}
(\byear{1970}).
\btitle{Efficient portfolio selection with quadratic and cubic utility}.
\bjournal{The Journal of Business}
\bvolume{43}
\bpages{181--189}.
\bptok{imsref}%
\end{barticle}
\endbibitem

\bibitem[\protect\citeauthoryear{Huang and Litzenberger}{1988}]{HuaLit88}
\begin{bbook}[mr]
\bauthor{\bsnm{Huang},~\bfnm{Chi-Fu}\binits{C.-F.}} \AND
  \bauthor{\bsnm{Litzenberger},~\bfnm{Robert~H.}\binits{R.~H.}}
(\byear{1988}).
\btitle{Foundations for Financial Economics}.
\bpublisher{North-Holland}, \blocation{New York}.
\bid{mr={0996240}}
\bptok{imsref}%
\end{bbook}
\endbibitem

\bibitem[\protect\citeauthoryear{Ingersoll}{1987}]{Ing}
\begin{bbook}[auto:STB|2013/03/04|13:35:07]
\bauthor{\bsnm{Ingersoll},~\bfnm{J.~E.}\binits{J.~E.}}
(\byear{1987}).
\btitle{Theory of Financial Decision Making}.
\bpublisher{Roman and Littlefield Publishers Inc.}, \blocation{Savage,
MD}.
\bptok{imsref}%
\end{bbook}
\endbibitem

\bibitem[\protect\citeauthoryear{Johnstone}{2012}]{Joh12}
\begin{barticle}[mr]
\bauthor{\bsnm{Johnstone},~\bfnm{D.~J.}\binits{D.~J.}}
(\byear{2012}).
\btitle{Log-optimal economic evaluation of probability forecasts}.
\bjournal{J. Roy. Statist. Soc. Ser. A}
\bvolume{175}
\bpages{661--689}.
\bid{doi={10.1111/j.1467-985X.2011.01011.x}, issn={0964-1998}, mr={2948369}}
\bptnote{check related}%
\bptok{imsref}%
\end{barticle}
\endbibitem

\bibitem[\protect\citeauthoryear{Johnstone and Lindley}{2011}]{JohLin11}
\begin{barticle}[mr]
\bauthor{\bsnm{Johnstone},~\bfnm{D.~J.}\binits{D.~J.}} \AND
  \bauthor{\bsnm{Lindley},~\bfnm{D.~V.}\binits{D.~V.}}
(\byear{2011}).
\btitle{Elementary proof that mean--variance implies quadratic utility}.
\bjournal{Theory and Decision}
\bvolume{70}
\bpages{149--155}.
\bid{doi={10.1007/s11238-010-9194-7}, issn={0040-5833}, mr={2753395}}
\bptok{imsref}%
\end{barticle}
\endbibitem

\bibitem[\protect\citeauthoryear{Lengwiler}{2004}]{Len04}
\begin{bbook}[auto:STB|2013/03/04|13:35:07]
\bauthor{\bsnm{Lengwiler},~\bfnm{Y.}\binits{Y.}}
(\byear{2004}).
\btitle{Microfoundations of Financial Economics: An Introduction to General
  Equilibrium Asset Pricing}.
\bpublisher{Princeton Univ. Press}, \blocation{Princeton}.
\bptok{imsref}%
\end{bbook}
\endbibitem

\bibitem[\protect\citeauthoryear{Levy}{1974}]{Lev74}
\begin{barticle}[auto:STB|2013/03/04|13:35:07]
\bauthor{\bsnm{Levy},~\bfnm{H.}\binits{H.}}
(\byear{1974}).
\btitle{The rationale of the mean--standard deviation analysis: Comment}.
\bjournal{American Economic Review}
\bvolume{64}
\bpages{434--442}.
\bptok{imsref}%
\end{barticle}
\endbibitem

\bibitem[\protect\citeauthoryear{Levy}{2006}]{Lev06}
\begin{bbook}[mr]
\bauthor{\bsnm{Levy},~\bfnm{Haim}\binits{H.}}
(\byear{2006}).
\btitle{Stochastic Dominance:
Investment Decision Making Under Uncertainty},
\bedition{2nd} ed.
\bseries{Studies in Risk and Uncertainty}
\bvolume{12}.
\bpublisher{Springer}, \blocation{New York}.
\bid{mr={2239375}}
\bptnote{check year}%
\bptok{imsref}%
\end{bbook}
\endbibitem

\bibitem[\protect\citeauthoryear{Levy}{2012}]{Lev12}
\begin{bbook}[auto:STB|2013/03/04|13:35:07]
\bauthor{\bsnm{Levy},~\bfnm{H.}\binits{H.}}
(\byear{2012}).
\btitle{The Capital Asset Pricing Model in the 21st Century: Analytical,
  Empirical, and Behavioral Perspectives}.
\bpublisher{Cambridge Univ. Press}, \blocation{Cambridge}.
\bptok{imsref}%
\end{bbook}
\endbibitem

\bibitem[\protect\citeauthoryear{Levy and Sarnat}{1969}]{LevSar69}
\begin{barticle}[auto:STB|2013/03/04|13:35:07]
\bauthor{\bsnm{Levy},~\bfnm{H.}\binits{H.}} \AND
  \bauthor{\bsnm{Sarnat},~\bfnm{M.}\binits{M.}}
(\byear{1969}).
\btitle{A note on indifference curves and uncertainty}.
\bjournal{The Swedish Journal of Economics}
\bvolume{71}
\bpages{206--208}.
\bptok{imsref}%
\end{barticle}
\endbibitem

\bibitem[\protect\citeauthoryear{Levy and Sarnat}{1972}]{LevSar72}
\begin{bbook}[auto:STB|2013/03/04|13:35:07]
\bauthor{\bsnm{Levy},~\bfnm{H.}\binits{H.}} \AND
  \bauthor{\bsnm{Sarnat},~\bfnm{M.}\binits{M.}}
(\byear{1972}).
\btitle{Investment and Portfolio Analysis}.
\bpublisher{Wiley}, \blocation{New York}.
\bptok{imsref}%
\end{bbook}
\endbibitem

\bibitem[\protect\citeauthoryear{Lintner}{1965}]{Lin65}
\begin{barticle}[auto:STB|2013/03/04|13:35:07]
\bauthor{\bsnm{Lintner},~\bfnm{J.}\binits{J.}}
(\byear{1965}).
\btitle{The valuation of risk assets and the selection of risky investments in
  stock portfolios and capital budgets}.
\bjournal{Rev. Econom. Statist.}
\bvolume{47}
\bpages{13--37}.
\bptok{imsref}%
\end{barticle}
\endbibitem

\bibitem[\protect\citeauthoryear{Liu}{2004}]{Liu04}
\begin{barticle}[mr]
\bauthor{\bsnm{Liu},~\bfnm{Liping}\binits{L.}}
(\byear{2004}).
\btitle{A new foundation for the mean--variance analysis}.
\bjournal{European J. Oper. Res.}
\bvolume{158}
\bpages{229--242}.
\bid{doi={10.1016/S0377-2217(03)00301-1}, issn={0377-2217}, mr={2063614}}
\bptok{imsref}%
\end{barticle}
\endbibitem

\bibitem[\protect\citeauthoryear{MacLean, Ziemba and Li}{2005}]{MacZieLi05}
\begin{barticle}[mr]
\bauthor{\bsnm{MacLean},~\bfnm{Leonard~C.}\binits{L.~C.}},
  \bauthor{\bsnm{Ziemba},~\bfnm{William~T.}\binits{W.~T.}} \AND
  \bauthor{\bsnm{Li},~\bfnm{Yuming}\binits{Y.}}
(\byear{2005}).
\btitle{Time to wealth goals in capital accumulation}.
\bjournal{Quant. Finance}
\bvolume{5}
\bpages{343--355}.
\bid{doi={10.1080/14697680500149552}, issn={1469-7688}, mr={2239384}}
\bptok{imsref}%
\end{barticle}
\endbibitem


\bibitem[\protect\citeauthoryear{Markowitz}{1952}]{Mar52}
\begin{barticle}[auto:STB|2013/03/04|13:35:07]
\bauthor{\bsnm{Markowitz},~\bfnm{H.~M.}\binits{H.~M.}}
(\byear{1952}).
\btitle{Portfolio selection}.
\bjournal{J. Finance}
\bvolume{7}
\bpages{77--91}.
\bptok{imsref}%
\end{barticle}
\endbibitem

\bibitem[\protect\citeauthoryear{Markowitz}{1959}]{Mar59}
\begin{bbook}[mr]
\bauthor{\bsnm{Markowitz},~\bfnm{Harry~M.}\binits{H.~M.}}
(\byear{1959}).
\btitle{Portfolio Selection: {E}fficient Diversification of Investments}.
\bseries{Cowles Foundation for Research in Economics at Yale University,
  Monograph}
\bvolume{16}.
\bpublisher{Wiley}, \blocation{New York}.
\bid{mr={0103768}}
\bptok{imsref}%
\end{bbook}
\endbibitem

\bibitem[\protect\citeauthoryear{Markowitz}{1991}]{Mar91}
\begin{barticle}[auto:STB|2013/03/04|13:35:07]
\bauthor{\bsnm{Markowitz},~\bfnm{H.~M.}\binits{H.~M.}}
(\byear{1991}).
\btitle{Foundations of portfolio theory}.
\bjournal{J.~Finance}
\bvolume{46}
\bpages{469--477}.
\bptok{imsref}%
\end{barticle}
\endbibitem

\bibitem[\protect\citeauthoryear{Markowitz}{2006}]{Mar}
\begin{barticle}[auto:STB|2013/03/04|13:35:07]
\bauthor{\bsnm{Markowitz},~\bfnm{H.~M.}\binits{H.~M.}}
(\byear{2006}).
\btitle{de Finetti scoops Markowitz}.
\bjournal{Journal of Investment Management}
\bvolume{4}
\bpages{5--18}.
\bptok{imsref}%
\end{barticle}
\endbibitem

\bibitem[\protect\citeauthoryear{Meyer}{1977}]{Mey77}
\begin{barticle}[mr]
\bauthor{\bsnm{Meyer},~\bfnm{Jack}\binits{J.}}
(\byear{1977}).
\btitle{Choice among distributions}.
\bjournal{J. Econom. Theory}
\bvolume{14}
\bpages{326--336}.
\bid{issn={0022-0531}, mr={0469189}}
\bptok{imsref}%
\end{barticle}
\endbibitem

\bibitem[\protect\citeauthoryear{Meyer}{1987}]{Mey87}
\begin{barticle}[auto:STB|2013/03/04|13:35:07]
\bauthor{\bsnm{Meyer},~\bfnm{J.}\binits{J.}}
(\byear{1987}).
\btitle{Two-moment decision models and expected utility}.
\bjournal{American Economic Review}
\bvolume{77}
\bpages{421--430}.
\bptok{imsref}%
\end{barticle}
\endbibitem

\bibitem[\protect\citeauthoryear{Mossin}{1966}]{Mos66}
\begin{barticle}[auto:STB|2013/04/24|11:25:54]
\bauthor{\bsnm{Mossin},~\bfnm{J.}\binits{J.}}
(\byear{1966}).
\btitle{Equilibrium in a capital asset market}.
\bjournal{Econometrica}
\bvolume{34}
\bpages{768--783}.
\bptok{imsref}%
\end{barticle}
\endbibitem

\bibitem[\protect\citeauthoryear{Mossin}{1973}]{autokey44}
\begin{bbook}[auto:STB|2013/03/04|13:35:07]
\bauthor{\bsnm{Mossin},~\bfnm{J.}\binits{J.}}
(\byear{1973}).
\btitle{Theory of Financial Markets}.
\bpublisher{Prentice Hall}, \blocation{Englewood Cliffs, NJ}.
\bptok{imsref}%
\end{bbook}
\endbibitem

\bibitem[\protect\citeauthoryear{Pennacchi}{2008}]{Pen08}
\begin{bbook}[auto:STB|2013/03/04|13:35:07]
\bauthor{\bsnm{Pennacchi},~\bfnm{G.}\binits{G.}}
(\byear{2008}).
\btitle{Theory of Asset Pricing}.
\bpublisher{Pearson}, \blocation{Boston}.
\bptok{imsref}%
\end{bbook}
\endbibitem

\bibitem[\protect\citeauthoryear{Pratt, Raiffa and
  Schlaifer}{1995}]{PraRaiSch95}
\begin{bbook}[mr]
\bauthor{\bsnm{Pratt},~\bfnm{John~W.}\binits{J.~W.}},
  \bauthor{\bsnm{Raiffa},~\bfnm{Howard}\binits{H.}} \AND
  \bauthor{\bsnm{Schlaifer},~\bfnm{Robert}\binits{R.}}
(\byear{1995}).
\btitle{Introduction to Statistical Decision Theory},
\bedition{2nd} ed.
\bpublisher{MIT Press}, \blocation{Cambridge, MA}.
\bptok{imsref}%
\end{bbook}
\endbibitem

\bibitem[\protect\citeauthoryear{Pressacco and Serafini}{2007}]{PreSer07}
\begin{barticle}[mr]
\bauthor{\bsnm{Pressacco},~\bfnm{Flavio}\binits{F.}} \AND
  \bauthor{\bsnm{Serafini},~\bfnm{Paolo}\binits{P.}}
(\byear{2007}).
\btitle{The origins of the mean--variance approach in finance: Revisiting de
  {F}inetti 65 years later}.
\bjournal{Decis. Econ. Finance}
\bvolume{30}
\bpages{19--49}.
\bid{doi={10.1007/s10203-007-0067-7}, issn={1593-8883}, mr={2323257}}
\bptok{imsref}%
\end{barticle}
\endbibitem

\bibitem[\protect\citeauthoryear{Rubinstein}{2006a}]{Rub06N1}
\begin{barticle}[auto:STB|2013/03/04|13:35:07]
\bauthor{\bsnm{Rubinstein},~\bfnm{M.}\binits{M.}}
(\byear{2006}a).
\btitle{Bruno de Finetti and mean--variance portfolio selection}.
\bjournal{Journal of Investment Management}
\bvolume{4}
\bpages{3--4}.
\bptok{imsref}%
\end{barticle}
\endbibitem

\bibitem[\protect\citeauthoryear{Rubinstein}{2006b}]{Rub06N2}
\begin{bbook}[auto:STB|2013/03/04|13:35:07]
\bauthor{\bsnm{Rubinstein},~\bfnm{M.}\binits{M.}}
(\byear{2006}b).
\btitle{A History of the Theory of Investments: My Annotated Bibliography}.
\bpublisher{Wiley}, \blocation{Hoboken, NJ}.
\bptok{imsref}%
\end{bbook}
\endbibitem

\bibitem[\protect\citeauthoryear{Sarnat}{1974}]{Sar74}
\begin{barticle}[auto:STB|2013/03/04|13:35:07]
\bauthor{\bsnm{Sarnat},~\bfnm{M.}\binits{M.}}
(\byear{1974}).
\btitle{A note on the implications of quadratic utility for portfolio theory}.
\bjournal{Journal of Financial and Quantitative Analysis}
\bvolume{9}
\bpages{687--689}.
\bptok{imsref}%
\end{barticle}
\endbibitem

\bibitem[\protect\citeauthoryear{Savage}{1954}]{Sav54}
\begin{bbook}[mr]
\bauthor{\bsnm{Savage},~\bfnm{Leonard~J.}\binits{L.~J.}}
(\byear{1954}).
\btitle{The Foundations of Statistics}.
\bpublisher{Wiley}, \blocation{New York}.
\bid{mr={0063582}}
\bptok{imsref}%
\end{bbook}
\endbibitem

\bibitem[\protect\citeauthoryear{Sharpe}{1964}]{Sha64}
\begin{barticle}[auto:STB|2013/03/04|13:35:07]
\bauthor{\bsnm{Sharpe},~\bfnm{W.~F.}\binits{W.~F.}}
(\byear{1964}).
\btitle{Capital asset prices: A theory of market equilibrium under conditions
  of risk}.
\bjournal{J. Finance}
\bvolume{19}
\bpages{425--442}.
\bptok{imsref}%
\end{barticle}
\endbibitem

\bibitem[\protect\citeauthoryear{Sharpe}{2007}]{Sha07}
\begin{barticle}[auto:STB|2013/03/04|13:35:07]
\bauthor{\bsnm{Sharpe},~\bfnm{W.~F.}\binits{W.~F.}}
(\byear{2007}).
\btitle{Expected utility asset allocation}.
\bjournal{Financial Analysts Journal}
\bvolume{63}
\bpages{18--30}.
\bptok{imsref}%
\end{barticle}
\endbibitem

\bibitem[\protect\citeauthoryear{Tobin}{1969}]{Tob69}
\begin{barticle}[auto:STB|2013/03/04|13:35:07]
\bauthor{\bsnm{Tobin},~\bfnm{J.}\binits{J.}}
(\byear{1969}).
\btitle{Comment on Borch and Feldstein}.
\bjournal{Rev. Econom. Stud.}
\bvolume{36}
\bpages{13--14}.
\bptok{imsref}%
\end{barticle}
\endbibitem

\bibitem[\protect\citeauthoryear{Von Neumann and Morgenstern}{1953}]{vonMor53}
\begin{bbook}[auto:STB|2013/03/04|13:35:07]
\bauthor{\bparticle{Von} \bsnm{Neumann},~\bfnm{J.}\binits{J.}} \AND
  \bauthor{\bsnm{Morgenstern},~\bfnm{O.}\binits{O.}}
(\byear{1953}).
\btitle{The Theory of Games and Economic Behavior},
\bedition{3rd} ed.
\bpublisher{Princeton Univ. Press}, \blocation{Princeton}.
\bptok{imsref}%
\end{bbook}
\endbibitem

\end{thebibliography}
\end{document}